\documentclass{vgtc}                          

\graphicspath{{figures/}{pictures/}{images/}{./}} 

\usepackage{times}                     

\usepackage{tabu}                      
\usepackage{booktabs}                  
\usepackage{lipsum}                    
\usepackage{mwe}                       

\usepackage{amsmath}
\usepackage{amssymb}
\usepackage{mathptmx}   
\usepackage{subcaption}

\vgtccategory{Research}

\vgtcinsertpkg

\title{Towards the Threshold: A Fall-Risk Anchored Pareto Framework for Virtual Reality Gait Feedback Selection for Individuals with Multiple Sclerosis}

\author{Nafisa Anjum\thanks{e-mail: nanjum1@students.kennesaw.edu}\\ %
        \scriptsize Kennesaw State University %
\and John Quarles\thanks{e-mail: john.quarles@utsa.edu}\\ %
     \scriptsize \centering University of Texas - San Antonio
     \and M. Rasel Mahmud \thanks{Corresponding author, e-mail: m.raselmahmud1@gmail.com}\\
     \scriptsize Kennesaw State University}

\abstract{
Reduced walking speed in people with multiple sclerosis (MS) is closely associated with a higher risk of falls. But VR rehabilitation studies often overlook whether improvements in performance come with acceptable cognitive and physical effort. This study introduces a \emph{threshold-anchored efficiency framework} that evaluates both performance and overall effort simultaneously. Here, a normative velocity target $(T=1.29 m/s)$ was established from the mean walking speed of non-fallers in an independent MS gait dataset, representing the speed at which MS patients walk when falls are not occurring. 34 adults with MS were then tested across eight VR feedback conditions (unimodal, bimodal and multimodal). Each condition was evaluated on how much it closed the velocity gap to $T$ and at what cognitive and physical burden cost. \emph{Pareto efficiency analysis} identified five non-dominated conditions : Static Visual, Spatial Auditory, Auditory+Visual, Auditory+Vibrotactile, and Multimodal -- while Spatial Vibrotactile and Vibrotactile+Visual were dominated. Spatial Auditory offered the best efficiency, closing 72.7\% of the gap at below-average burden. Multimodal closed the most gap (95.6\%) but imposed the highest burden. These findings give clinicians a principled basis for selecting VR feedback strategies matched to each patient's capacity.

} 

\keywords{Virtual Reality, Multiple Sclerosis, Feedback, Gait, Pareto Efficiency, Motor Rehabilitation.}

\begin{document}
\maketitle

\section{Introduction}

As one of the most basic human activities, walking depends on the smooth coordination of sensory information from the visual, auditory, and somatosensory systems. In VR, this coordination can be disturbed due to limitations of head-mounted displays (HMDs), such as a limited field of view, system latency, and altered motion cues \cite{chang2020virtual}. These issues can cause instability, slower walking speed, and changes in stride patterns, especially for people with mobility impairments like Multiple Sclerosis (MS) \cite{mahmud2022auditory,samaraweera2013latency, mahmud2023eyes}. Multiple sclerosis is a chronic neurological disorder that affects more than 2.8 million people worldwide \cite{walton2020ms}. Among its many symptoms, gait impairment is one of the most disabling and functionally consequential \cite{motl2012exercise}. Walking difficulties in MS are not merely a quality-of-life concern, they carry direct safety implications. Between 50 and 70 percent of people with MS experience at least one fall per year, and recurrent falls are associated with injury, fear of movement, and accelerated loss of independence \cite{nilsagard2009falls, sosnoff2011falls}. Among measurable gait parameters, walking speed is one of the most sensitive indicators of fall risk: slower walkers are consistently overrepresented among fallers in MS populations. As such, normative walking speed
benchmarks have been proposed as clinically actionable targets
for rehabilitation planning \cite{comber2017gait,kalron2014relationship}.

At the same time, VR is increasingly being used in healthcare for applications such as rehabilitation, motor training, and assistive therapy \cite{kim2023virtual,keshner2021untapped}. 
Unlike traditional training modalities, VR
allows precise layering of sensory cues -- visual, auditory, and vibrotactile, to guide movement correction during walking tasks \cite{ mirelman2010vr}. This flexibility makes VR particularly appealing for MS, where motor and sensory profiles vary substantially across individuals \cite{calabro2017role}. Studies have
demonstrated that VR-based gait training can improve walking speed, balance, and confidence in people with MS \cite{massetti2016virtual, feng2019virtual, de2016effect}.
Despite such progress, a persistent methodological gap exists in how VR feedback conditions are evaluated. The predominant approach asks a single question: which condition produces the best performance outcome? This framing overlooks an equally important clinical reality that feedback conditions differ not only in how much they help, but in how much they demand of
the user. Richer multimodal conditions may drive greater walking speed gains while simultaneously imposing higher cognitive and physical load. This trade-off matters especially for MS patients whose neurological reserves are already limited.
However, this issue has received limited systematic attention in the VR rehabilitation literature. 
Efficiency-based evaluation frameworks are well established in human-computer interaction (HCI) \cite{iso9241} but have rarely been applied to rehabilitation condition selection. More critically, prior work has evaluated VR conditions almost exclusively against within-study
performance comparisons \cite{mahmud2022standing, janeh2019gait, mahmud2022auditory}, rather than against external benchmarks that carry independent clinical meaning. This leaves open a fundamental question: \textit{do any feedback conditions
actually move patients toward a functionally safer state, and if so, which ones do so most efficiently?}

This study addresses these gaps by introducing a
threshold-anchored efficiency framework for VR gait feedback selection in MS. We derive a normative velocity target $(T)$ from the mean walking speed of non-fallers in an independent MS gait dataset \cite{meyer2022open} -- representing the speed at which MS patients walk when they are not falling. We then evaluate a combination of unimodal, bimodal and multimodal VR feedback conditions across 34 adults with MS on two criteria: how much each condition closes the velocity gap to $T$, and at what cost in mental load and fatigue. Pareto efficiency analysis \cite{deb2011multi} identifies conditions that cannot be simultaneously outperformed on both dimensions. This yields a principled, evidence-based menu for selecting feedback conditions in VR. The key contributions of this study are thus threefold:
\begin{itemize}
\item \textbf{Threshold-Anchored Benchmarking for VR Gait Evaluation:} We introduce a novel application of external clinical benchmarking in multimodal VR feedback research. Rather than comparing conditions against within-study means, we ground evaluation against an externally derived fall-risk velocity target, established from an independent MS gait dataset. 
\item \textbf{First Efficiency Profiling of Eight VR Feedback Modalities for Rarely Explored People with MS:} Recruiting 34 adults with MS related mobility impairment, we systematically evaluated unimodal, bimodal, and multimodal VR feedback conditions. We show that two widely used conditions, Spatial Vibrotactile and Vibrotactile+Visual, are dominated. Spatial Auditory feedback offers the most favorable performance-to-burden ratio among clinically viable options.
\item \textbf{Pareto Efficiency Analysis as a Condition Selection Tool:} We demonstrate that Pareto efficiency analysis can identify which VR feedback conditions advance MS patients toward a safe velocity target without disproportionate cognitive and physical burden, producing an actionable frontier of dominant conditions.
\end{itemize}

\section{Background and Related Work}\label{sec:background}
\subsection{Gait Impairment and Fall Risk in Multiple Sclerosis}
Gait impairment is one of the most prevalent and functionally limiting consequences of Multiple Sclerosis. Walking difficulties in MS arise from a combination of motor weakness, spasticity, sensory deficits, and fatigue, producing a characteristic pattern of reduced walking speed, shortened stride length, increased step variability, and asymmetric limb loading \cite{comber2017gait}.  Reductions in walking speed below normative benchmarks have been associated with increased fall frequency, reduced community participation and earlier transition to mobility aid dependence \cite{motl2012exercise, kalron2014relationship}. To prevent fall risk, walking speed is one of the most clinically sensitive gait indicators in MS \cite{d2012cognitive}. The Timed 25-Foot Walk test, one of the most widely used clinical gait assessments in MS, uses walking speed as its primary outcome precisely because of this sensitivity \cite{motl2017validity}. 
\subsection{VR for Gait Rehabilitation}
VR has emerged as a promising platform for gait rehabilitation across neurological populations, offering controlled, repeatable environments that can be enriched with augmented sensory feedback \cite{laver2017vr, massetti2016virtual}. Winter et al. \cite{winter2021immersive} investigated treadmill-based gait training under immersive, semi-immersive, and non-VR conditions and found that immersive VR significantly increased walking speed in individuals with MS and stroke. Another study \cite{janeh2019gait} evaluated a VR-based gait training method for patients with Parkinson’s disease that manipulated step length to improve gait symmetry. Their results showed improvements in step width and swing time, demonstrating the potential of VR as a rehabilitation tool for neurological conditions. Similarly, Guo et al. \cite{guo2015mobility} reported that individuals with mobility impairments exhibited different gait responses such as changes in walking velocity, step length, and stride length, compared to participants without impairments. However, VR immersion also introduces a well-documented gait disruption effect. Head-mounted displays restrict the visual field, introduce latency, and alter motion cues, producing visual-vestibular mismatch. Such mismatch, in turn, reduces walking speed, shortens stride length, and increases gait variability relative to real-world walking \cite{horsak2021overground, canessa2019comparing, mahmud2023auditory}. So, VR without augmented feedback may worsen gait, making feedback a structural necessity rather than an optional enhancement.

\subsection{Assistive Feedback in VR Walking}
The majority of VR gait feedback research has examined single sensory modalities in isolation. Lee et al. \cite{lee2015influence} explored the use of visual feedback to improve postural stability in stroke patients. In their study, a laser pointer attached to the participant’s head projected onto a board, allowing researchers to assess how well patients coordinated head movements with eye tracking. Results showed that this visual feedback approach improved balance during standing exercises. Mahmud et al. \cite{mahmud2023eyes} studied the effect of visual feedback on walking in VR. Participants stood on a balance board while wearing a VR headset and received different visual cues, including a head-fixed frame, a rhythmic visual frame, and a display that changed based on body sway. The results showed improvements in gait measures such as velocity, step length, and stride length, particularly for participants with mobility impairments compared to those without impairments. Auditory feedback, particularly spatialized cues, improves postural orientation and walking regularity by providing reliable spatial information that remains unaffected by visual-vestibular conflict \cite{gandemer2017spatial, mahmud2022auditory}. Weller et al. \cite{weller2022redirected} investigated how auditory cues, especially spatialized footstep sounds, can be used to guide redirected walking in VR. Furthermore, Kingma et al. \cite{kingma2019vibrotactile} examined the effects of vibrotactile feedback on balance and movement in individuals with severe bilateral vestibular dysfunction. In the study, 39 participants wore a vibrotactile belt with 12 tactors placed around the waist that delivered real-time feedback during balance tasks. The results showed a significant improvement in balance and mobility ($p < 0.001$). Vibrotactile feedback delivers real-time directional corrections through body-mounted tactors and has demonstrated significant balance and gait improvements in MS population \cite{mahmud2025vibrotactile}. 

Conceptual frameworks of multisensory integration propose that combining complementary sensory cues can reduce perceptual uncertainty and improve motor performance, especially in situations where sensory signals are conflicting \cite{roy2021multisensory}. Consistent with this, studies combining auditory and vibrotactile channels have reported larger gait improvements than either channel alone \cite{machado2023novel}. Leonardis et al. \cite{leonardis2014multisensory} investigated multisensory feedback that can be coupled with body movements to enhance the sense of embodiment in VR. Their study focused on how vestibular and proprioceptive signals together influence perceived embodiment during immersive walking. However, the relationship between feedback complexity and user experience is not simple. Complex multimodal conditions impose higher cognitive and physical load, and it remains unclear whether the added burden leads to proportional improvements in performance.
\subsection{Efficiency Frameworks in Rehabilitation and HCI}
The concept of efficiency, achieving the highest output for a given cost, is well established in human–computer interaction. It is considered one of the three main components of usability, alongside effectiveness and satisfaction \cite{iso9241}. In rehabilitation research, efficiency has often been discussed in terms of treatment intensity and patient tolerance \cite{kwakkel2004understanding}. However, formal frameworks that evaluate rehabilitation conditions using multiple criteria simultaneously are still rare. 

Pareto efficiency analysis, widely used in engineering and economics, is a method for evaluating solutions when multiple objectives must be considered at the same time \cite{deb2011multi}. A solution is considered \emph{Pareto-efficient} if no other option can improve one objective without worsening another. The set of these optimal trade-offs forms the \emph{Pareto frontier}, which represents the best available choices across competing objectives. Despite its broad use in optimization problems, Pareto analysis has not, to the best of our knowledge, been applied to evaluating VR rehabilitation conditions.  In this work, we introduce this approach to identify feedback conditions that achieve strong performance improvements while maintaining manageable user burden. Importantly, the method also reveals \emph{dominated} conditions that perform worse while imposing equal or greater burden, which should be excluded from clinical consideration.
The absence of such efficiency frameworks in VR rehabilitation research has practical consequences. Studies that focus only on performance improvements may recommend conditions that impose excessive cognitive or physical load on patients. This issue is particularly relevant for individuals with MS, where cognitive fatigue is a major barrier to rehabilitation participation and daily activity \cite{induruwa2012fatigue}. The \emph{threshold-anchored efficiency framework} proposed in this paper addresses this gap by evaluating VR feedback conditions based on both performance improvement and user burden, providing a more balanced and clinically meaningful basis for selecting rehabilitation strategies.
\section{Methodology}
\begin{figure}[t]
  \centering
\includegraphics[width=0.95\columnwidth]{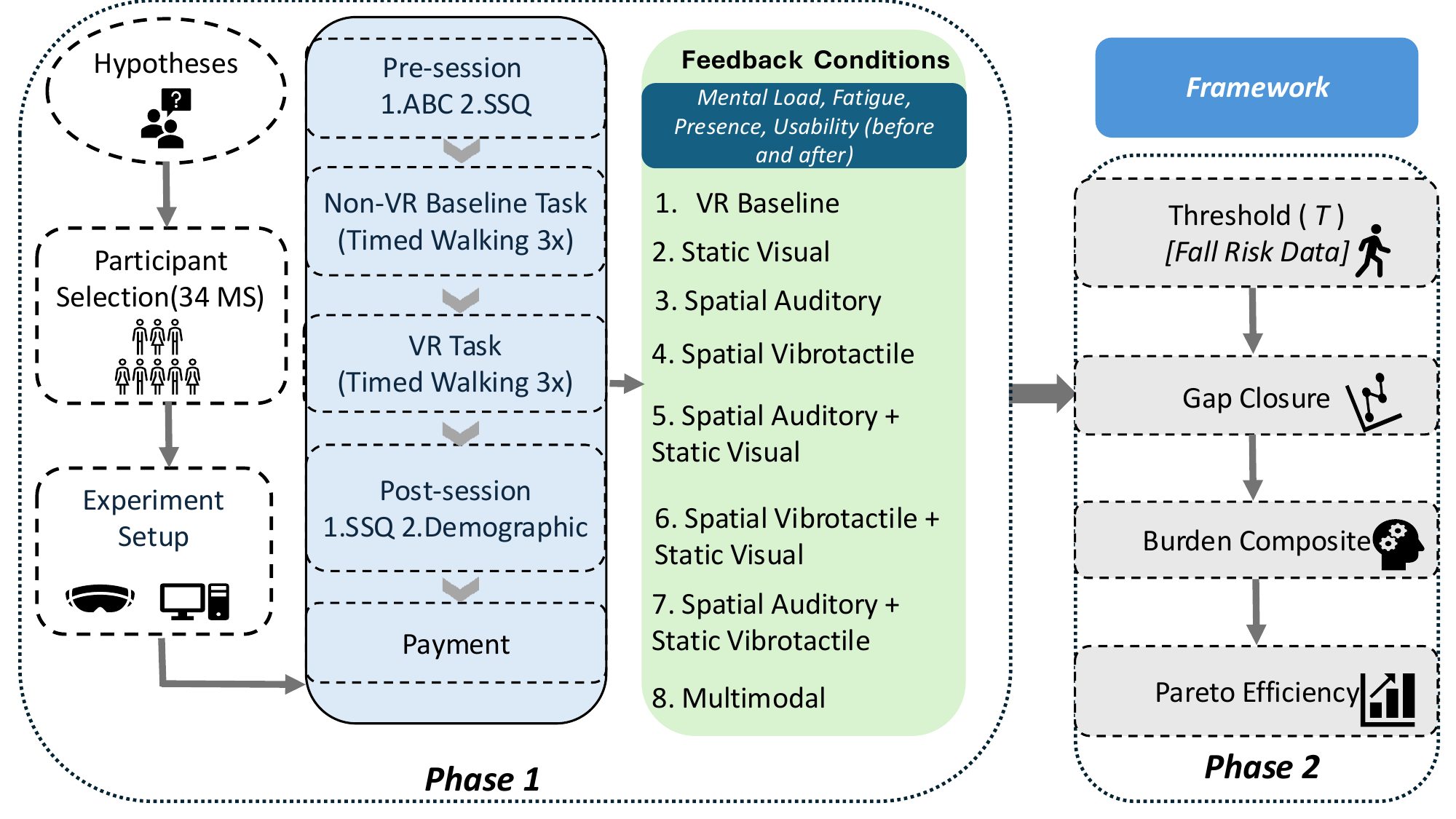}
  \caption{Methodology of the study.}
  \label{fig:method}
  \vspace{-0.2in}
\end{figure}

The study was approved by the Institutional Review Board (IRB). The entire methodology has been outlined in Fig. \ref{fig:method}.
\subsection{Hypotheses}
Building on the established destabilizing effects of VR immersion on gait as outlined in Section \ref{sec:background} and the gap identified in the literature regarding efficiency-based condition evaluation, we formulate the following hypotheses:
\begin{itemize}
\item \textbf{H1 -- VR Immersion Disrupts Gait in MS:} Walking velocity under VR Baseline will be significantly lower than under Non-VR Baseline in participants with MS. This effect has been consistently reported across neurological populations \cite{samaraweera2013latency,mahmud2023eyes} and is included here to confirm that the gait disruption motivating the feedback conditions is present in this specific MS sample.
\item \textbf{H2 -- Feedback Conditions Advance Participants Towards $T$:} At least a subset of VR feedback conditions will produce mean walking velocities significantly closer to the normative fall-risk threshold $T$ than VR Baseline. 
\item \textbf{H3 -- Performance Gain and Perceived Burden are Not Uniformly Correlated:} Velocity Gap closure and subjective burden will not increase proportionally across conditions, that is, a more effective condition will not always impose greater cost. This decoupling is the structural prerequisite that makes Pareto analysis necessary.
\item \textbf{H4 -- A Non-Trivial Pareto Frontier Exists:} A subset of the eight feedback conditions will form the Pareto frontier. At least one condition will be dominated, meaning another condition achieves the same or greater velocity gap closure with equal or lower burden. If such a frontier exists, it would show that the efficiency framework meaningfully distinguishes between conditions rather than simply ranking them by performance.
\item \textbf{H5 -- The Pareto Frontier is Robust to How Burden is Measured:} The set of frontier conditions identified using the combined burden measure will remain similar when mental load and fatigue are analyzed separately. If the frontier stays stable across these different measures, it would suggest that the results reflect real efficiency differences rather than being influenced by how burden was calculated.
\end{itemize}
\subsection{Participants}
34 adults with mobility impairment due to multiple sclerosis  were recruited from the local MS community under an approved IRB protocol. All participants provided written informed consent prior to enrollment. The group comprised of 17 males and 17 females with a mean age of 46.5 years, mean height of 164.84 cm, and mean weight of 82.79 kg (details in Table \ref{tab:participants}). Inclusion required a confirmed MS diagnosis and the ability to walk independently, with or without an assistive device. Participants with concurrent neurological or orthopedic conditions independently affecting gait were excluded. All sessions were conducted in a climate-controlled laboratory of approximately 600 square feet, with only the participant and research staff present.
\subsection{Experimental Setup}
\subsubsection{Apparatus and Virtual Environment}
Walking performance was captured using a GAITRite instrumented walkway system \cite{gaitrite_manual} -- a 12-foot pressure-sensitive mat recording spatiotemporal gait parameters including walking velocity, cadence, step length, stride length, step time, swing time, single and double support intervals, and base of support. Walking velocity was defined as:
$v=\frac{d}{t}$ where
$d$ is the total distance traveled in centimeters and  $t$ is the elapsed time in seconds. We also recorded additional gait metrics such as cadence, step length, step time, swing and stance phases, and single and double support intervals. But walking velocity served as the primary outcome measure and was converted to meters per second for all analyses.

The virtual environment was rendered through an HTC Vive Pro head-mounted display (110° field of view, 90 Hz refresh rate) with integrated noise-isolating headphones. The VR application was developed in Unity3D and run on a high-performance workstation (Windows 10, Intel Core i7 at 4.35 GHz, 32 GB RAM, NVIDIA GeForce RTX 2080). GAITRite data were synchronized with Unity3D through a custom socket-based integration, ensuring real-time alignment between physical walking and the virtual simulation. Because real walking tasks may increase fall risk, participants were secured using a full-body suspension harness system (Kaye Products Inc.). The setup consisted of a body harness, thigh cuffs, and an overhead support frame that prevented falls while allowing natural movement. This safety arrangement enabled participants to walk comfortably while remaining fully protected, providing a controlled and safe experimental environment.

\subsubsection{Multimodal Feedback System}
\begin{figure}[h!]
\centering
\includegraphics[width=6.25cm,height=2.5cm]{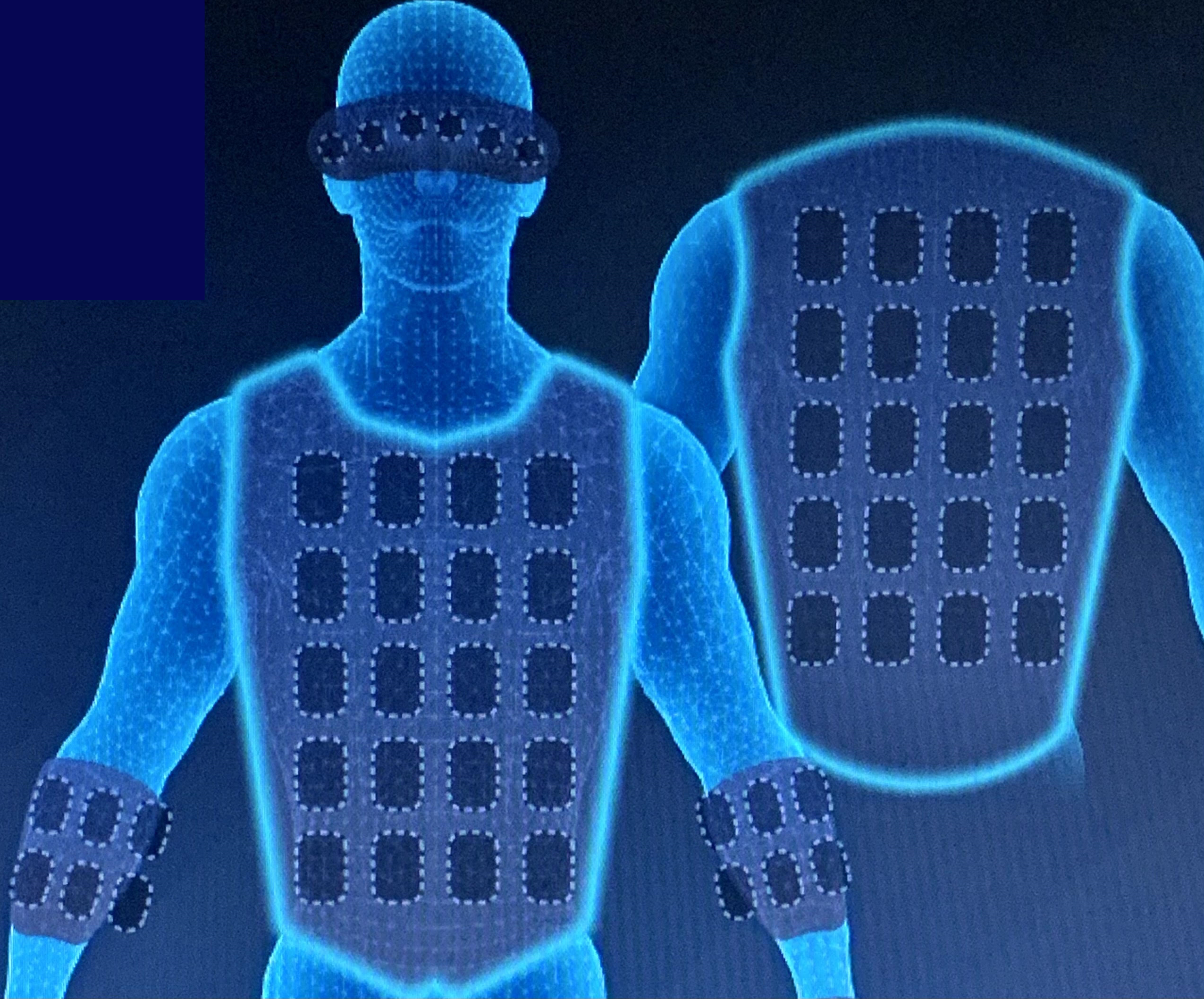}
\caption{ Vibration motors positions.}
\label{fig:vibration motors}
\vspace{-0.1in}
\end{figure}

To provide feedback through multiple sensory channels, bHaptics hardware\footnote{\url{https://www.bhaptics.com/}} was used. Participants wore a wireless vest containing 40 vibrotactile actuators (20 on the front and 20 on the back) that delivered distributed tactile feedback across the torso. The vest weighed 3.7 pounds and included adjustable shoulder fasteners. Participants also wore forearm sleeves equipped with six vibrotactile motors on each arm, providing localized tactile cues. Each sleeve weighed approximately 0.66 pounds. In addition, a lightweight forehead device containing six vibrotactile motors was attached to the HMD, allowing targeted tactile stimulation near the head. This device weighed approximately 0.18 pounds. The placement of all vibration motors is shown in Fig.~\ref{fig:vibration motors}. Along with tactile feedback, spatialized audio cues were generated using Unity’s Resonance Audio toolkit, and static visual anchors were placed within the virtual scene to provide stable reference points. Together, these components created a trimodal feedback system integrating auditory, tactile, and visual inputs.

\begin{figure}[h!]
\centering
\includegraphics[width=0.85\columnwidth,height=4cm]{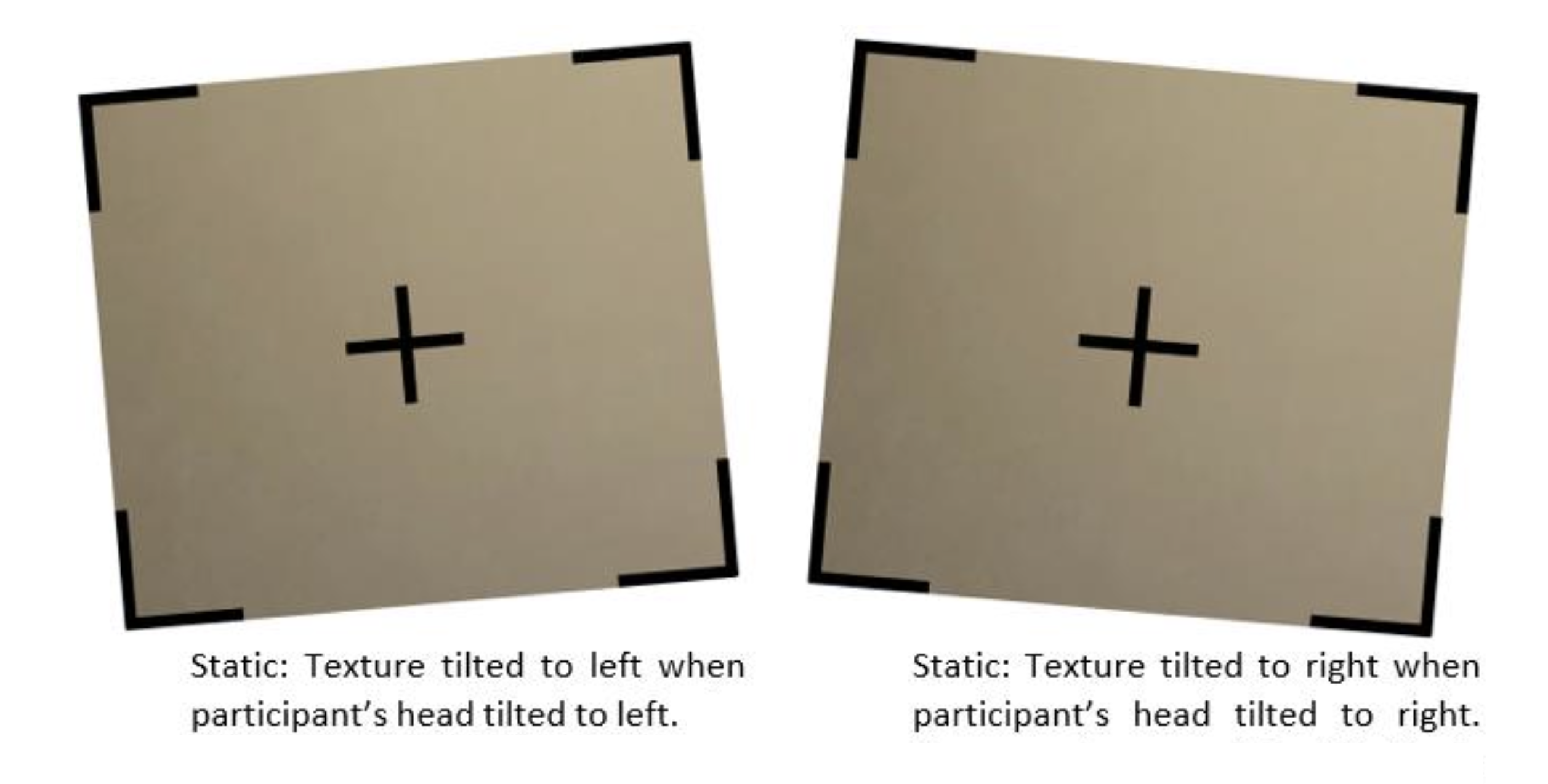}
\vspace{-0.1in}
\caption{Static Visual Cues.}
\label{fig:visual_cues}
\vspace{-0.2in}
\end{figure}

\begin{table}[t]
\caption{Participant demographics.}
\label{tab:participants}
\centering
\resizebox{\columnwidth}{!}{%
\begin{tabular}{lcccccccc}
\hline
\textbf{Group} & \multicolumn{2}{c}{\textbf{Participants}} & \multicolumn{2}{c}{\textbf{Age (years)}} & \multicolumn{2}{c}{\textbf{Height (cm)}} & \multicolumn{2}{c}{\textbf{Weight (kg)}} \\
 & Male & Female & Mean & SD & Mean & SD & Mean & SD \\
\hline
MS & 17 & 17 & 46.5 & 13.0 & 164.84 & 12.62 & 82.79 & 22.18 \\
\hline
\end{tabular}%
}
\vspace{-0.2in}
\end{table}

\subsection{Experimental Feedback Conditions}

The feedback conditions used in the study are summarized in Table~\ref{tab:conditions}. Each condition was designed to examine how different sensory cues influence walking performance in VR.

\subsubsection{Auditory Feedback}
Auditory guidance was implemented using spatial audio generated with the Google Resonance Audio SDK, which has been shown to support gait regulation in immersive VR environments \cite{mahmud2022auditory, mahmud2023auditory,  mahmud2024multimodal}. Sounds were delivered through over-ear headphones integrated with the HMD, enabling accurate three-dimensional sound localization. White noise was used instead of music or tones because it has been reported to improve motor control through stochastic resonance effects \cite{helps2014different}. Unlike natural sounds such as footsteps or environmental audio, white noise does not carry semantic meaning or rhythmic structure, which helps reduce potential dual-task interference during walking. 

\subsubsection{Vibrotactile Feedback}
Vibrotactile cues were provided to deliver spatial tactile guidance during walking, as previous studies have demonstrated their effectiveness for improving motor performance \cite{mahmud2025vibrotactile, mahmud2022standing}. Eccentric Rotating Mass (ERM) tactors embedded in the vest, arm sleeves, and forehead array generated vibrations linked to head orientation and body position. Head rotations controlled the vibration patterns in the forehead array, while spatial position was mapped to tactors placed on the torso. The Unity environment generated spatial audio signals that were converted into vibration patterns using the bHaptics Audio-to-Vibrotactile engine, ensuring that tactile feedback remained synchronized with auditory cues. The vibration intensity was initially set to 50\% and adjusted if needed for participant comfort.

\subsubsection{Visual Feedback}
Visual guidance was provided using static visual anchors placed within the participant’s field of view (Fig.~\ref{fig:visual_cues}). The anchor consisted of five compact frames, including four L-shaped corner markers and a central crosshair. It was attached to the front wall and remained head-fixed, meaning it moved with the participant’s head rather than remaining fixed in the virtual world. This design, adapted from prior VR locomotion research \cite{shahnewaz2021static}, provided a stable visual reference intended to reduce disorientation and support consistent stride alignment.

\subsubsection{Vibrotactile + Visual Feedback}
This condition combined tactile and visual cues. Participants experienced spatial vibrotactile signals along with the static visual anchor simultaneously. The goal was to examine whether combining body-based tactile feedback with a stable visual reference could improve walking regularity and reduce gait deviations compared to using either cue alone.

\subsubsection{Auditory + Visual Feedback}
In this condition, spatial audio cues were presented together with the visual anchor. Participants received directional auditory guidance while also viewing the stabilizing visual frame. 

\subsubsection{Auditory + Vibrotactile Feedback}
This condition delivered synchronized auditory and tactile signals. Participants received spatial audio along with corresponding vibrotactile feedback, allowing them to perceive directional guidance through both hearing and touch. The purpose of this setup was to evaluate whether multisensory reinforcement could provide stronger corrective cues for walking stability than either modality alone.

\subsubsection{Multimodal Feedback}
The multimodal condition combined auditory, vibrotactile, and visual feedback simultaneously. This condition was designed to explore the full potential of multisensory integration for improving walking performance in VR. Measures such as walking speed, stride consistency, and trajectory deviation were analyzed to determine whether trimodal feedback offered advantages over unimodal or bimodal conditions.

\subsubsection{No Feedback (Non-VR and VR Baseline)}
Two baseline conditions were included for comparison. In the VR baseline, participants walked in the virtual environment while wearing the headset and equipment but received no auditory, vibrotactile, or visual feedback cues. This condition represented natural walking performance in VR without assistance. In the non-VR baseline, participants walked normally without the VR headset or feedback devices, providing a reference for natural walking performance outside the virtual environment.
\begin{table}[t]
\caption{Experimental feedback conditions used during walking in VR.}
\label{tab:conditions}
\centering
\renewcommand{\arraystretch}{1.2}
\begin{tabular}{p{3.2cm} p{5cm}}
\hline
\textbf{Condition} & \textbf{Feedback} \\
\hline
Auditory & Spatialized white-noise cues via headphones. \\

Vibrotactile & Tactile signals delivered through vest, sleeves, and forehead actuators. \\

Visual & Static visual anchors to support orientation. \\

Auditory + Visual & Combined auditory cues and visual anchors. \\

Vibrotactile + Visual & Tactile signals paired with visual anchors. \\

Auditory + Vibrotactile & Synchronized auditory and tactile guidance. \\

Multimodal & Auditory, vibrotactile, and visual feedback combined. \\

VR Baseline & Walking in VR with no feedback cues. \\

Non-VR Baseline & Walking without VR or feedback devices. \\
\hline
\end{tabular}
\vspace{-0.2in}
\end{table}

\subsection{Subjective Burden Assessment}

Subjective burden was evaluated using two measures collected before and after each experimental condition. \textbf{Mental load} was rated on a scale from 0 (no mental effort) to 10 (extremely high effort), capturing the perceived cognitive demand associated with each feedback condition. \textbf{Fatigue} was rated on the same 0--10 scale to reflect perceived physical tiredness. Only post-condition ratings were used in the analysis.
A burden composite was created by first computing grand z-scores for each measure across all participants and conditions, and then averaging the standardized scores:

\begin{equation}
B_c = \frac{z(ML_c) + z(F_c)}{2}
\end{equation}
where $z(ML_c)$ and $z(F_c)$ represent the grand z-scored mental load and fatigue ratings for condition $c$, respectively.

Grand z-scores were calculated as:

\begin{equation}
z(x_{ic}) = \frac{x_{ic} - \bar{\mu}}{\bar{\sigma}}
\end{equation}

where $\bar{\mu}$ and $\bar{\sigma}$ denote the grand mean and standard deviation of the corresponding measure computed across all participants and conditions.

Internal consistency of the composite measure was evaluated using Cronbach's alpha ($\alpha = 0.86$), indicating good reliability and supporting the use of equal weighting when combining the two measures.

\subsection{Normative Velocity Threshold}\label{sec:normative}
\subsubsection{Reference Dataset}
The normative velocity threshold $T$ was obtained from a publicly available MS fall risk gait dataset \cite{meyer2022open}. This dataset contains gait and clinical information collected from community-dwelling adults with multiple sclerosis (MS) during both laboratory and home walking sessions. For the present study, only the laboratory session data were used, as they closely match the controlled walking conditions used in our experiment. The dataset included gait recordings from 41 participants with MS, along with a binary label indicating whether each participant had experienced a fall in the previous year.

\subsubsection{Fall Label Extraction and Participant Matching}

Fall history labels were obtained from the clinical data file accompanying the gait recordings. Participants were categorized as \emph{fallers} ($n = 22$) if they reported at least one fall in the past year, and \emph{non-fallers} ($n = 19$) otherwise. These labels were matched with gait measurements using participant IDs, resulting in a complete dataset containing both fall status and gait parameters for all 41 participants.

\subsubsection{Group-Level Velocity Analysis}
\begin{figure}[t]
  \centering
  \includegraphics[width=0.85\columnwidth, height=5cm]{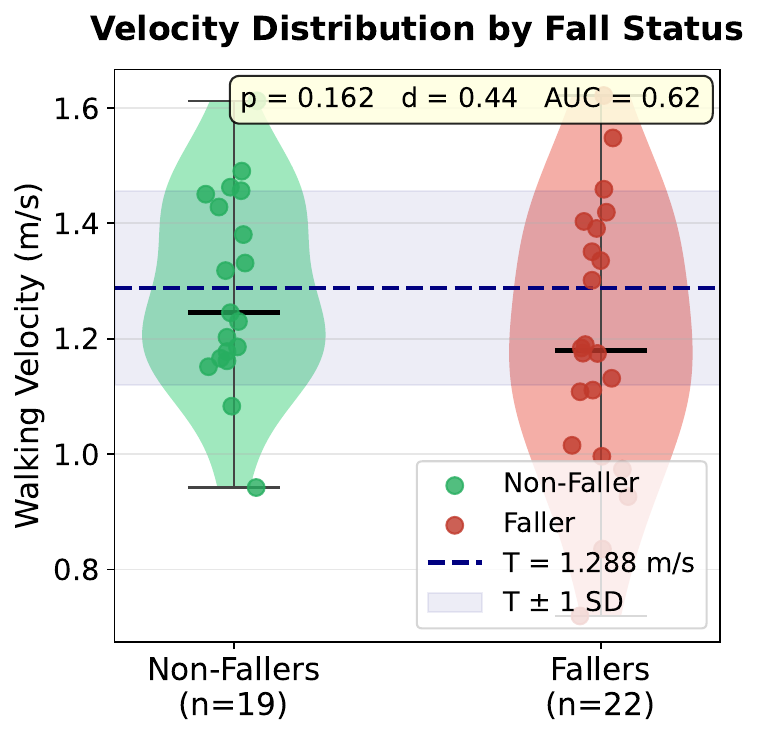}
  \caption{Velocity Distribution in MS Fall Risk Data.}
  \label{fig:velocity violin}
  \vspace{-0.2in}
\end{figure}

Average walking velocity was calculated separately for fallers and non-fallers using the laboratory session data (Fig. \ref{fig:velocity violin}). The descriptive statistics were:
\begin{itemize}
    \item \textbf{Non-fallers:} $\bar{v} = 1.29$ m/s, $SD = 0.17$ m/s ($n = 19$)
    \item \textbf{Fallers:} $\bar{v} = 1.20$ m/s, $SD = 0.23$ m/s ($n = 22$)
\end{itemize}
The difference between the two groups was $\Delta \bar{v} = 0.09$ m/s with a small effect size (Cohen's $d = 0.44$). An independent samples $t$-test produced $p = .16$, while a Mann-Whitney $U$ test produced $p = .19$. These results indicate that the difference was not statistically significant, which is reasonable given the relatively small sample size and the variability in gait characteristics among individuals with MS.

\subsubsection{Evaluation of a Diagnostic Threshold}
Before defining the threshold, we examined whether walking velocity could be used to classify fallers and non-fallers through Receiver Operating Characteristic (ROC) analysis. The ROC results were:
\begin{equation}
AUC = 0.62, \quad p = .16
\end{equation}
An AUC of 0.62 indicates weak discriminative ability, only slightly better than random chance. Because of this limited predictive power, using a ROC-derived cut-off as a diagnostic threshold would likely lead to high misclassification rates. Therefore, a diagnostic interpretation was not adopted.

\subsubsection{Threshold Definition}
Instead, a normative benchmarking approach was used. The threshold $T$ was defined as the mean walking velocity of the non-faller group in the reference dataset:
\vspace{-0.1in}
\begin{equation}
T = \bar{v}_{\text{non-faller}} = 1.29 \text{ m/s} \quad (SD = 0.17 \text{ m/s})
\end{equation}

A lower bound was also estimated as one standard deviation below this mean:
\vspace{-0.1in}
\begin{equation}
T_{\text{lower}} = T - SD = 1.12 \text{ m/s}
\end{equation}
This lower bound was used only for sensitivity checks, while $T$ served as the primary reference value in the gap closure analysis. Importantly, $T$ is not treated as a clinical diagnostic cut-point or a guaranteed fall-prevention threshold. Instead, it represents the walking speed typically observed in MS individuals who do not fall. Therefore, improvements toward $T$ are interpreted as movement toward a safer functional walking speed, without implying that reaching this value directly prevents falls.
\subsection{Gap Closure Computation}

For each feedback condition $c$ in our data, the velocity gap relative to the threshold was computed as:
\vspace{-0.1in}
\begin{equation}
\text{gap}(c) = T - \bar{v}_c
\end{equation}

where $\bar{v}_c$ represents the mean walking velocity across all 34 MS participants under condition $c$.

Gap closure relative to the VR Baseline condition was defined:
\vspace{-0.1in}
\begin{equation}
\text{closure}(c) = \text{gap}(c_{\text{VRB}}) - \text{gap}(c)
\end{equation}

A positive closure value indicates that condition $c$ reduces the velocity gap to $T$ more than the VR baseline without feedback. The percentage of the gap closed was calculated as:

\begin{equation}
\text{closed}(c) =
\frac{\text{closure}(c)}{\left|\text{gap}(c_{\text{VRB}})\right|}
\times 100
\end{equation}

The Non-VR Baseline condition was excluded from efficiency analysis because it does not represent a VR feedback design option.

\subsection{Pareto Efficiency Analysis}

Pareto efficiency analysis was used to identify VR feedback conditions that cannot be simultaneously outperformed in both performance and burden. A condition $c_i$ is considered Pareto-dominated if another condition $c_j$ satisfies with at least one strict inequality:

\begin{equation}
\text{closure}(c_j) \geq \text{closure}(c_i)
\quad \text{and} \quad
B_{c_j} \leq B_{c_i}
\end{equation}
Conditions that are not dominated by any other candidate form the \emph{Pareto frontier}, representing feedback options that offer meaningful trade-offs between performance improvement and burden. The VR Baseline condition was excluded from frontier candidacy because it serves as the reference point ($\text{closure} = 0$ by definition). To summarize the overall value of each frontier condition in a single interpretable quantity, an efficiency ratio was computed as:
\vspace{-0.1in}
\begin{equation}
\text{Efficiency Ratio}(c) = \frac{\text{closure}(c)}{B_c + 2.0}
\end{equation}

where the constant offset of $2.0$ is added to the denominator to ensure all values remain positive, since the burden composite ranges approximately from $-1.6$ to $+1.4$ across conditions. A higher efficiency ratio indicates greater gap closure per unit of burden cost. This metric is used to rank frontier conditions and is reported alongside Pareto status in Sec. \ref{sec: results}. To evaluate robustness, sensitivity analyses were conducted by repeating the Pareto analysis using mental load alone and fatigue alone as the burden measure. 
\subsection{Secondary Measures}

In addition to the primary performance and burden metrics, several secondary measures were collected to provide additional context on participant experience and baseline characteristics.

\textbf{Presence.} Participants’ sense of immersion in the virtual environment was assessed using a single-item presence (SIP) rating ranging from 0 (not present at all) to 10 (fully present), following the approach in \cite{bouchard2004reliability}. This measure captured how engaging the VR walking environment felt under different feedback conditions.

\textbf{Usability.} System usability was evaluated using the question “Is the system usable?” rated on a Likert scale from fully agree to fully disagree. This measure provided insight into the perceived practicality of the feedback conditions.

\textbf{Balance Confidence.} The Activities-Specific Balance Confidence (ABC) Scale \cite{powell1995activities} was used to assess participants’ confidence in performing everyday physical activities. Scores were categorized as high ($>80\%$), moderate (50--80\%), or low ($<50\%$), providing context about participants’ perceived balance ability.

\textbf{Cybersickness.} The Simulator Sickness Questionnaire (SSQ) \cite{kennedy1993simulator} was administered before and after the study to monitor symptoms such as nausea, oculomotor strain, and disorientation that could affect VR walking performance.
\subsection{Study Procedure}

\begin{figure}[t]
\centering
\begin{subfigure}{0.48\columnwidth}
    \centering
    \includegraphics[width=\linewidth, angle=270]{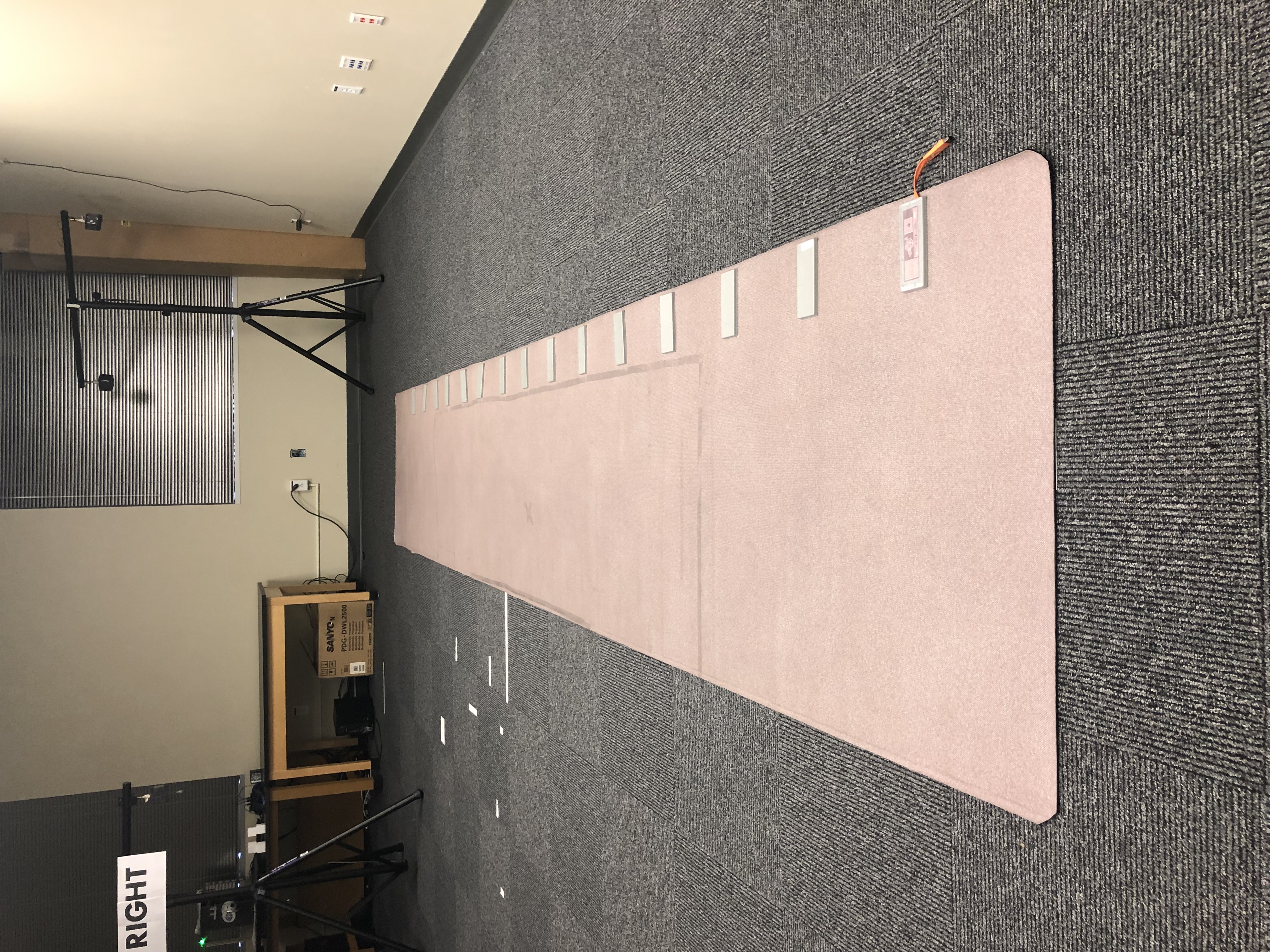}
    \caption{Real environment}
\end{subfigure}
\hfill
\begin{subfigure}{0.38\columnwidth}
    \centering
    \includegraphics[width=\linewidth, height=4.1cm]{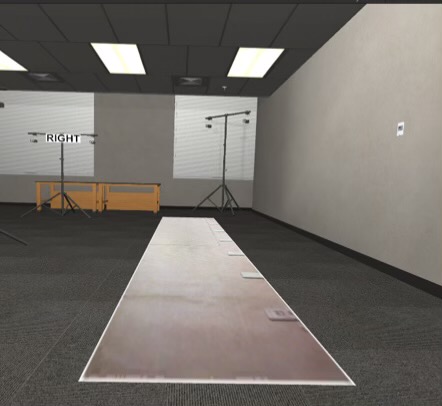}
    \caption{Virtual environment}
\end{subfigure}
\caption{Comparison between real and virtual environments for the timed walking task.}
\label{fig:gait_real_virtual}
\vspace{-0.2in}
\end{figure}

\textbf{Preparation and Safety Protocols.}
All participants provided written informed consent before participation and were given a full explanation of the study procedures. At the start of the session, participants completed two baseline questionnaires, the ABC Scale and the SSQ, and removed footwear that could interfere with GAITRite measurements. Feedback settings were individually calibrated for comfort, including audio volume, vibration intensity, and visual anchor thickness. A comparison between real-world and virtual walking environments is shown in Fig.~\ref{fig:gait_real_virtual}. 

\begin{figure}[ht!]
	\centering
\includegraphics[width=0.8\linewidth]{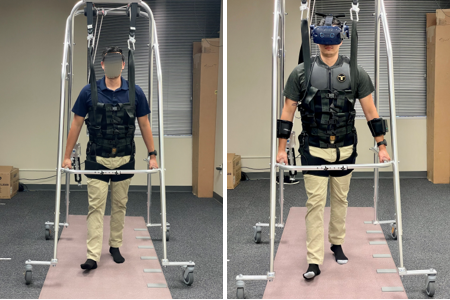}
	\caption{Participants used harnesses while performing the timed walking task utilizing the GAITRite system in both the real world (left) and virtual environment (right).}
    \label{fig:walking}
    \vspace{-0.2in}
\end{figure}

\textbf{Walking Task.}
Walking performance was first assessed in the real-world environment. Participants wore a full-body safety harness connected to an overhead suspension system and completed three timed walking trials. They were instructed to walk at a comfortable pace across the GAITRite walkway and perform 180-degree turns at each end. Because the GAITRite does not record turning movements, participants stepped off the mat during turns before continuing the trial. During the tasks, ambulation time (via stopwatch) and gait velocity (via GAITRite) were recorded. The same task was then repeated in the VR environment. Depending on the experimental condition, participants received auditory, vibrotactile, visual, dual-modality, or multimodal feedback. A no-feedback VR baseline condition was also included. These were presented in counterbalanced order to minimize learning and fatigue effects. Three trials were recorded per condition. An example of the task in real (left) and virtual (right) environment has been shown in Fig.~\ref{fig:walking}. Subjective ratings of mental load, fatigue, presence, and usability were collected before and after each condition. Upon completing all conditions, participants filled out a post-session SSQ and demographic questionnaire. They were compensated \$30 per hour along with a parking validation.
  
\section{Results} \label{sec: results}
\subsection{Statistical Analysis}
\begin{figure}[t]
  \centering
  \includegraphics[width=0.85\columnwidth]{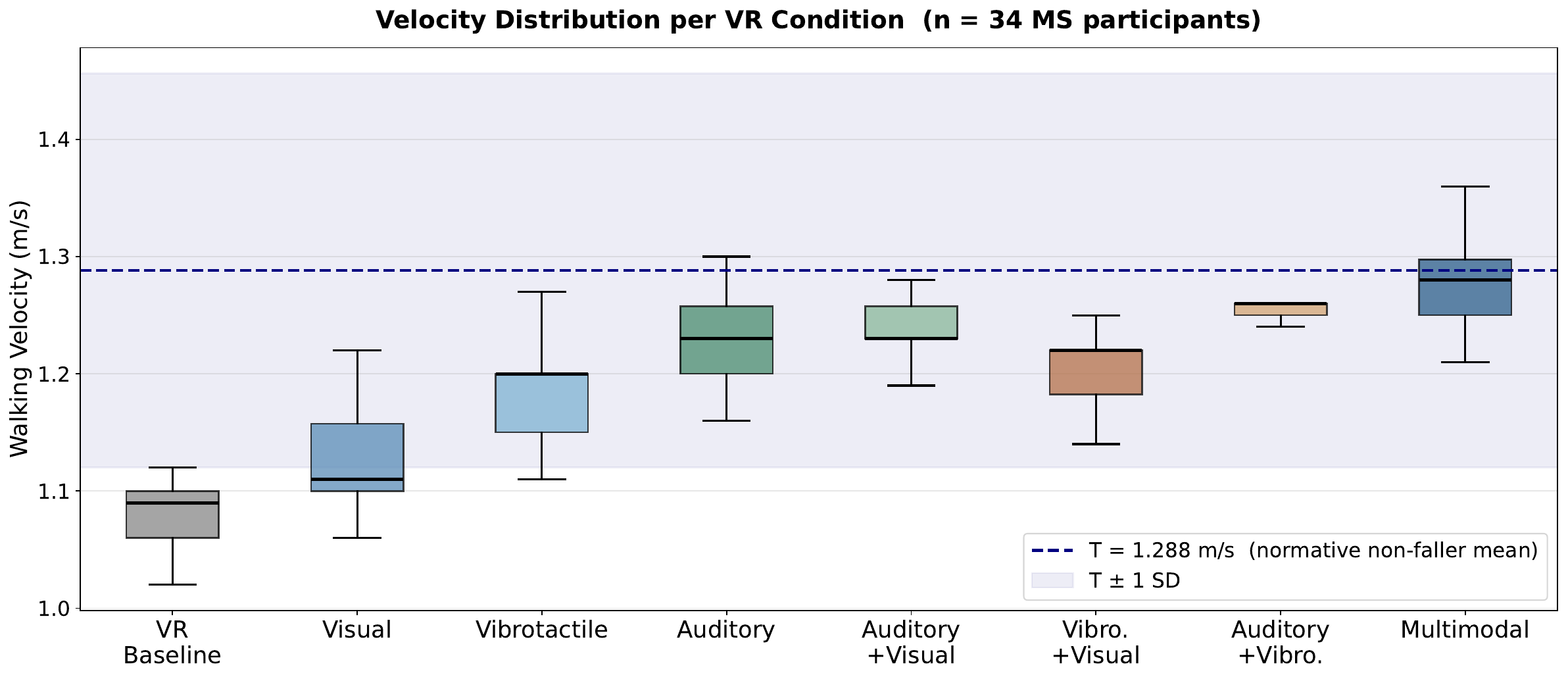}
  \caption{Walking Velocity of MS participants.}
  \label{fig:walking MS}
  \vspace{-0.1in}
\end{figure}
Walking velocity data for the MS participant group ($n = 34$) were tested for normality using the \textbf{Shapiro--Wilk} test. Visual inspection of histograms, Q--Q plots, and boxplots was also performed to support the normality assessment. The results indicated that velocity values were normally distributed across all nine experimental conditions ($p > .05$). A one-way repeated-measures \textbf{ANOVA} with \textit{Condition} as the within-subjects factor (9 levels) revealed a significant main effect on walking velocity ($F(8, 264) = 171.30$, $p < .001$, $\eta_p^2 = 0.839$), indicating that condition membership explained 83.9\% of the variance. 

To further examine differences between conditions, multiple pairwise comparisons were performed using post-hoc paired sample \textbf{$t$-tests} for all condition pairs of our MS data (detailed in Table \ref{tab:posthoc_ttest_mims}). For subjective measures, non-parametric tests were used due to the ordinal nature of Likert-scale responses. Specifically, Friedman tests were conducted to detect overall differences across conditions, followed by Wilcoxon signed-rank tests for pairwise comparisons.

\subsection{VR-Induced Gait Disruption}

Walking velocity under the VR Baseline condition was significantly lower than under the Non-VR Baseline condition ($\bar{v}_{\text{Non-VR}} = 1.2926$ m/s vs. $\bar{v}_{\text{VRB}} = 1.0797$ m/s; $t(33) = 21.30$, $p < .001$, $d = 3.65$). This corresponds to a reduction of 0.2129 m/s due to VR immersion alone, before any feedback was introduced. The effect size ($d = 3.65$) is exceptionally large, indicating that VR-induced gait disruption is substantial in this MS sample. Compared with the Non-VR Baseline condition, the VR Baseline condition showed significant decreases in cadence, step length, and stride length, along with significant increases in step time and swing time ($p < .001$ for all comparisons). The observed VR-induced velocity reduction of 0.2129 m/s also establishes the reference gap used in subsequent gap-closure analyses. Specifically, it represents the difference between unassisted walking speed in VR and the normative fall-risk threshold ($T = 1.29$ m/s).
\subsection{Normative Threshold Derivation}
The normative threshold $T = 1.29$ m/s was derived from the reference dataset as described in Sec. \ref{sec:normative}. The the non-faller mean was used as a directional benchmark rather than a diagnostic cut-point due to weak group separation ($d = 0.44$, $AUC = 0.62$, $p = .16$).
\subsection{Walking Velocity and Gap Closure}
\begin{figure}[t]
  \centering
\includegraphics[width=0.85\columnwidth]{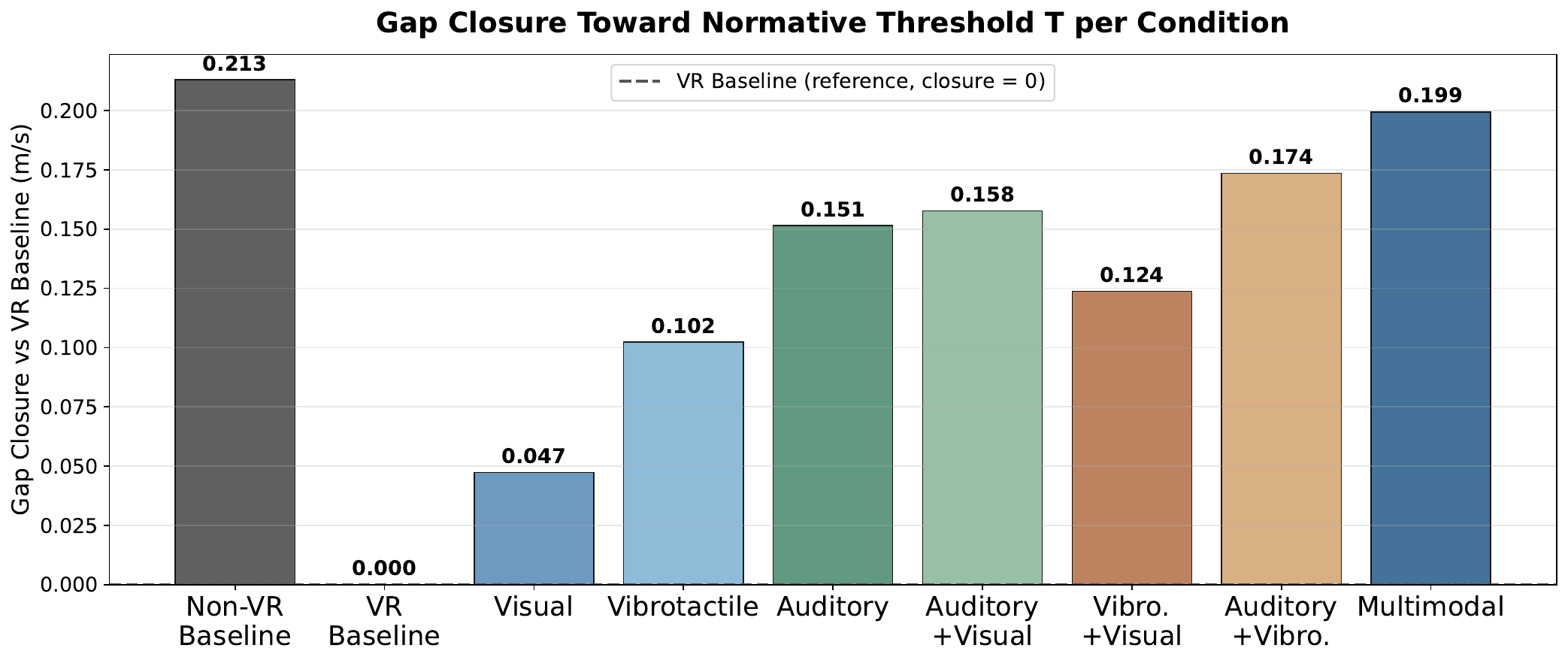}
  \caption{Gap Closure Analysis.}
  \label{fig:gap_closure}
  \vspace{-0.2in}
\end{figure}

\begin{table}[t]
\caption{Within-group pairwise comparisons of study conditions for MS participants. 
($t =$ t-statistic, $p =$ probability value where * indicates significance at $p<.05$, 
$d =$ Cohen's $d$ effect size).}
\label{tab:posthoc_ttest_mims}
\centering
\begin{tabular}{lccc}
\toprule
Study Conditions & $t(33)$ & $p$ & $d$ \\
\midrule
VR Baseline vs. Non-VR Baseline     & 21.30 & $<$.001* & 3.65 \\
VR Baseline vs. Static Visual       & 7.03  & $<$.001* & 1.21 \\
VR Baseline vs. Spatial Vibro. & 13.05 & $<$.001* & 2.24 \\
VR Baseline vs. Spatial Auditory    & 20.44 & $<$.001* & 3.51 \\
VR Baseline vs. Multimodal          & 23.59 & $<$.001* & 4.04 \\
Spatial Vibro. vs. Static Visual & 7.17 & $<$.001* & 1.23 \\
Spatial Vibro. vs. Spatial Auditory & 5.84 & $<$.001* & 1.00 \\
Auditory vs. Vibro.+Visual & 3.92 & 0.015 & 0.67 \\
Auditory vs. Auditory+Visual     & 1.64 & $>$.999 & 0.28 \\
Auditory+Visual vs. Auditory+Vibro. & 5.21 & $<$.001* & 0.89 \\
Multimodal vs. Auditory+Vibro.       & 3.99 & $<$.001* & 0.68 \\
\bottomrule
\end{tabular}

\end{table}

All eight VR feedback conditions resulted in higher walking velocities than the VR Baseline (Fig. \ref{fig:walking MS}). Pairwise comparisons confirmed significant improvements for seven of the eight conditions. Several cross-modal comparisons further highlight the structure of the efficiency hierarchy. The comparison between Auditory + Visual and Auditory + Vibrotactile yielded $t(33) = 5.21$, $p < .001$, $d = 0.89$, indicating a large effect size between these adjacent frontier conditions. Spatial Vibrotactile vs.\ Static Visual yielded $t(33) = 7.17$, $p < .001$, $d = 1.23$, and Spatial Vibrotactile vs.\ Spatial Auditory yielded $t(33) = 5.84$, $p < .001$, $d = 1.00$. We also observed that multimodal and bimodal conditions led to significant increases in cadence, step length, and stride length, along with significant decreases in step time and swing time ($p < .001$), compared to unimodal conditions.These results confirm that the velocity differences underlying the gap closure rankings are statistically robust across the condition hierarchy.

Table~\ref{tab:gapclosure} reveals a wide range of gap closure across conditions, from 22.7\%
(Static Visual) to 95.6\% (Multimodal). So, the choice of feedback modality significantly influences how closely MS participants approach the threshold $T$. The Non-VR Baseline slightly exceeded the threshold $T$ by 0.004 m/s ($\bar{v} = 1.2926$ m/s), confirming that $T$ lies within the achievable walking velocity range for this MS sample. This indicates that the threshold represents a realistic functional target rather than an aspirational benchmark. Critically, conditions that rank similarly on raw velocity do not rank similarly on gap closure relative to $T$ (Fig.\ref{fig:gap_closure}), motivating the efficiency analysis in Sec. \ref{sec:efficiency analysis}.

\begin{table}[t]
\centering
\caption{Walking velocity (m/s), gap to $T$ (m/s), and gap closure (m/s) per condition. $\star$ denotes Pareto-dominant conditions.}
\label{tab:gapclosure}
\footnotesize
\setlength{\tabcolsep}{3pt}
\begin{tabular}{lcccc}
\hline
Condition & $\bar{v}$ & Gap to $T$ & Closure & \% Closed \\
\hline
Non-VR Baseline              & 1.2926 & -0.0044 & 0.2129 & 102.1 \\
Multimodal $\star$           & 1.2791 & 0.0091  & 0.1994 & 95.6  \\
Auditory+Vibrotactile $\star$& 1.2532 & 0.0350  & 0.1735 & 83.2  \\
Auditory+Visual $\star$      & 1.2374 & 0.0508  & 0.1577 & 75.6  \\
Spatial Auditory $\star$     & 1.2312 & 0.0570  & 0.1515 & 72.7  \\
Vibrotactile+Visual          & 1.2035 & 0.0847  & 0.1238 & 59.4  \\
Spatial Vibrotactile         & 1.1821 & 0.1061  & 0.1024 & 49.1  \\
Static Visual $\star$        & 1.1271 & 0.1611  & 0.0474 & 22.7  \\
VR Baseline                  & 1.0797 & 0.2085  & 0.0000 & 0.0   \\
\hline
\end{tabular}
\vspace{-0.2in}
\end{table}

\subsection{Subjective Burden Across Conditions}

Table~\ref{tab:burden} reports the mean mental load, fatigue, and burden composite for each condition. \textit{Cronbach's alpha} confirmed good internal consistency between mental load and fatigue before combining the measures ($\alpha = 0.86$), supporting the use of an equal-weight composite score. The burden values increased with feedback complexity, beginning with the VR Baseline ($B = -1.592$), followed by unimodal and bimodal conditions, and reaching the highest value in the Multimodal condition ($B = +1.407$).
\begin{table}[t]
\centering
\caption{Subjective burden per condition ($\alpha = 0.86$). Mental load and fatigue are mean scores on a 0--10 scale.}
\label{tab:burden}
\footnotesize
\setlength{\tabcolsep}{3pt}
\begin{tabular}{lccc}
\hline
Condition & Mental & Fatigue & Burden Composite\\
\hline
VR Baseline           & 1.00 & 1.00 & -1.592 \\
Static Visual         & 2.56 & 2.56 & -0.572 \\
Spatial Auditory      & 3.03 & 3.09 & -0.246 \\
Spatial Vibrotactile  & 3.24 & 3.44 & -0.064 \\
Auditory+Visual       & 3.50 & 3.59 & 0.072 \\
Vibrotactile+Visual   & 3.59 & 3.85 & 0.185 \\
Auditory+Vibrotactile & 4.53 & 4.82 & 0.810 \\
Multimodal            & 5.44 & 5.74 & 1.407 \\
\hline
\end{tabular}
\vspace{-0.1in}
\end{table}
Pairwise comparisons of the raw mental load and fatigue scores showed that the Multimodal condition produced significantly higher mental load ($p = .04$) and fatigue ($p = .01$) compared to unimodal conditions. No other pairwise differences reached statistical significance, suggesting that the burden differences among the remaining conditions followed a gradual trend rather than distinct step changes.

Presence ratings did not differ significantly across conditions ($p > .05$). In addition, all participants reported that every condition was usable, resulting in no detectable differences in usability ratings. With a  p-value of $.06$, and a an effect size of $(d = 0.04)$, participants were not considerably affected by cybersickness as observed from the SSQ.

\subsection{Pareto Efficiency Analysis} \label{sec:efficiency analysis}
Before applying Pareto analysis, a Friedman test confirmed that gap closure values differed significantly across the eight VR feedback conditions ($\chi^2(7) = 205.17$, $p < .001$), indicating that the choice of feedback condition has a statistically robust effect on how closely MS participants approach the threshold $T$. Table~\ref{tab:pareto} reports gap closure, burden composite, Pareto status, and efficiency ratio for the seven VR feedback conditions (VR Baseline excluded). Five conditions formed the Pareto frontier, while two were dominated.

\begin{table}[t]
\centering
\caption{Pareto efficiency results for VR feedback conditions. $\star$ = Pareto-dominant, $\times$ = dominated.}
\label{tab:pareto}
\footnotesize
\setlength{\tabcolsep}{3pt}
\begin{tabular}{lcccc}
\hline
Condition & Closure & Burden & Status & Eff. \\
\hline
Multimodal            & 0.1994 & 1.407 & $\star$ & 0.059 \\
Auditory+Vibrotactile & 0.1735 & 0.810 & $\star$ & 0.062 \\
Auditory+Visual       & 0.1577 & 0.072 & $\star$ & 0.076 \\
Spatial Auditory      & 0.1515 & -0.246 & $\star$ & 0.086 \\
Vibrotactile+Visual   & 0.1238 & 0.185 & $\times$ & 0.057 \\
Spatial Vibrotactile  & 0.1024 & -0.064 & $\times$ & 0.053 \\
Static Visual         & 0.0474 & -0.572 & $\star$ & 0.033 \\
\hline
\end{tabular}
\vspace{-0.2in}
\end{table}

\textbf{Vibrotactile+Visual} was dominated by Auditory+Visual, which closed 75.6\% of the velocity gap compared to 59.4\% at lower burden ($B=0.072$ vs.\ $0.185$). Similarly, \textbf{Spatial Vibrotactile} was dominated by Spatial Auditory, which closed 72.7\% of the gap compared to 49.1\% while also requiring lower burden ($B=-0.246$ vs.\ $-0.064$). Among the frontier conditions, \textbf{Spatial Auditory} achieved the highest efficiency ratio (0.086), indicating the greatest gap closure per unit of burden. In contrast, \textbf{Multimodal} achieved the largest absolute improvement in walking velocity (0.1994 m/s, corresponding to 95.6\% gap closure) but also imposed the highest burden ($B = 1.407$).

\subsection{Sensitivity Analysis}
Table~\ref{tab:sensitivity} shows the Pareto frontier obtained using three different burden measures: the composite burden score, mental load alone, and fatigue alone. The frontier remained identical across all three cases. The same five conditions were Pareto-dominant and the same two were dominated in every analysis, demonstrating that the efficiency structure is robust to how user burden is defined.

\begin{table}[t]
\centering
\caption{Pareto frontier stability across burden operationalization.\\
\footnotesize Y = Pareto-dominant; N = dominated by at least one condition. Identical Y/N patterns across columns indicate robustness to the choice of burden measure.}
\label{tab:sensitivity}
\footnotesize
\setlength{\tabcolsep}{4pt}
\begin{tabular}{lccc}
\hline
Condition & Composite & Mental Load Only & Fatigue Only \\
\hline
Multimodal            & Y & Y & Y \\
Auditory+Vibrotactile & Y & Y & Y \\
Auditory+Visual       & Y & Y & Y \\
Spatial Auditory      & Y & Y & Y \\
Static Visual         & Y & Y & Y \\
Vibrotactile+Visual   & N & N & N \\
Spatial Vibrotactile  & N & N & N \\
\hline
\end{tabular}
\vspace{-0.2in}
\end{table}

\section{Discussion}
\subsection{The Cost of Immersion}
The reduction in walking velocity under the VR Baseline condition relative to the Non-VR Baseline was approximately 0.21 m/s ($d = 3.65$), representing one of the largest effects observed in this study. Similar gait disruptions in VR have been reported in both healthy adults \cite{horsak2021overground, canessa2019comparing} and neurological populations \cite{ferdous2018investigating}. In our data, the mean velocity under VR Baseline was 1.08 m/s, which is 0.21 m/s below the normative threshold ($T = 1.29$ m/s). This indicates that unassisted VR walking places participants in a velocity range associated with increased fall risk, supporting Hypothesis \textbf{H1} and motivating the need for feedback-based interventions.
These findings suggest that VR walking without feedback may reduce gait performance in MS. Therefore, sensory feedback mechanisms should be considered an essential component of VR-based gait rehabilitation.
\subsection{Advancing Toward the Threshold: The Clinical Case for VR Feedback in MS}
All eight VR feedback conditions reduced the velocity gap to the threshold $T$ compared with the VR Baseline, with gap closure ranging from 22.7\% (Static Visual) to 95.6\% (Multimodal). This finding supports Hypothesis \textbf{H2} and demonstrates that VR feedback does more than simply improve performance within the virtual environment. Instead, it moves participants toward a walking velocity profile associated with non-fallers in an independent dataset. This distinction is important because improvements relative to an independent reference point may be more meaningful than improvements measured only within the experimental environment.

The Multimodal condition produced the largest improvement, closing 95.6\% of the gap and yielding a mean walking velocity of $\bar{v} = 1.279$ m/s. This value lies within 0.01 m/s of the threshold $T$, which is within the measurement noise of the GAITRite system \cite{gaitrite_manual}. Such improvements are consistent with multisensory integration theory, which suggests that combining multiple sensory signals can reduce perceptual uncertainty and improve motor performance \cite{roy2021multisensory}. In VR environments where visual reliability may be reduced, auditory and vibrotactile cues can provide stable body-referenced information \cite{wall2010application, gandemer2017spatial}. This likely explains why conditions containing auditory or vibrotactile feedback consistently outperformed purely visual feedback. Similar patterns have also been reported in VR balance studies \cite{mahmud2022auditory} and broader multimodal feedback research \cite{sigrist2013augmented}.

Although Static Visual feedback produced the smallest improvement (22.7\%), it still moved participants toward the threshold $T$ while imposing the lowest burden among feedback conditions. This suggests that lower-intensity feedback may be appropriate in early rehabilitation stages when patient capacity is limited.
\subsection{Performance Versus Burden: Why Effectiveness Alone Is Not Enough}

A key finding of this study is that performance improvement and subjective burden do not increase together across VR feedback conditions, supporting Hypothesis \textbf{H3}. For example, the Spatial Auditory condition closed 72.7\% of the velocity gap while maintaining below-average burden ($B = -0.25$). In contrast, Vibrotactile+Visual closed only 59.4\% of the gap while imposing higher burden ($B = +0.19$). This means that Vibrotactile+Visual both improves performance less and costs more effort than Spatial Auditory. Without an efficiency framework, a simple performance ranking could mistakenly favor the inferior condition.

In VR environments, visual and vestibular signals may become less reliable due to factors such as limited field of view and altered motion cues \cite{ferdous2018investigating}. One possible explanation, consistent with sensory reweighting theory \cite{asslander2014sensory}, is that the nervous system tends to rely more on sensory channels that remain reliable, such as auditory and vibrotactile signals. Spatial Auditory feedback likely benefits from this process because it provides clear directional information through a channel that remains stable during VR locomotion. In contrast, Vibrotactile+Visual combines a reliable tactile cue with a visual cue that may already be less trusted in VR. This may increase cognitive demand without producing proportional performance benefits.
These findings also challenge a common assumption in multimodal interface design -- that adding more sensory channels always improves performance \cite{sigrist2013augmented}. Our results suggest that the way sensory channels are combined is more important than the number of channels used. Effective feedback should align with the sensory signals that remain reliable during locomotion.

This issue is particularly important for individuals with MS, as cognitive fatigue affects up to 90\% of patients and can reduce participation in rehabilitation \cite{askari2021cognitive}. Therefore, burden should be treated as a key design consideration, and evaluating both performance and burden together provides a more meaningful basis for selecting VR feedback strategies.

\subsection{The Pareto Frontier as a Clinical Decision Tool}
\begin{figure}[t]
  \centering
  \includegraphics[width=0.75\columnwidth, height=4cm]{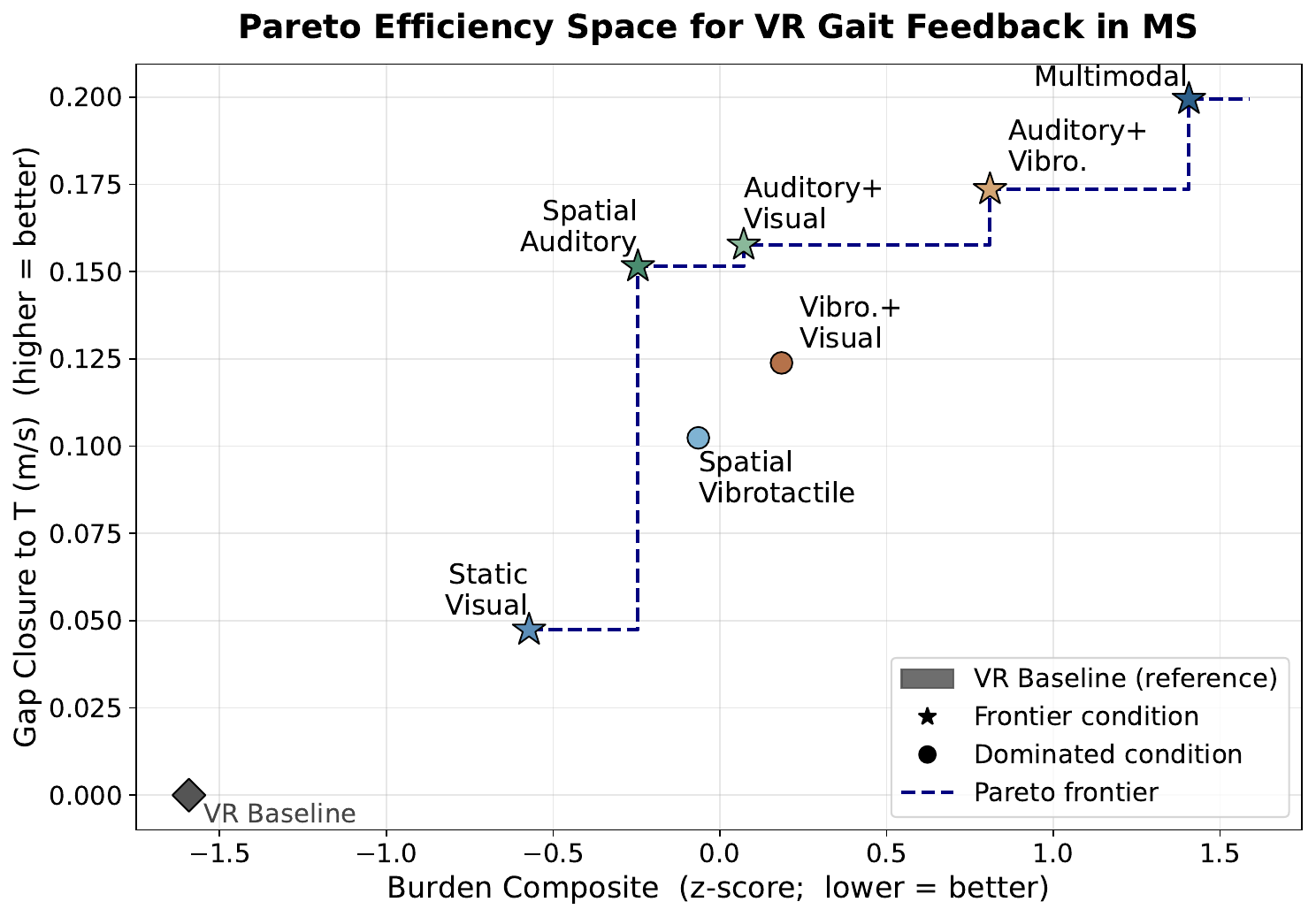}
  \caption{Pareto Efficiency Space.}
  \label{fig:pareto MS}
  \vspace{-0.2in}
\end{figure}

The Pareto analysis identified five dominant conditions and two dominated conditions, confirming \textbf{H4}, observed in the efficiency space of Fig. \ref{fig:pareto MS}. The dominated conditions--Spatial Vibrotactile and Vibrotactile+Visual, should generally not be recommended because another condition always provides equal or better performance at equal or lower burden. In other words, these conditions do not offer a meaningful advantage within the performance–burden trade-off space. The domination of Spatial Vibrotactile by Spatial Auditory highlights the importance of sensory modality. Both conditions provide spatial guidance through a single channel, but auditory cues produced higher gap closure and lower burden. This result is consistent with prior studies showing that spatial auditory signals are easily integrated with locomotor control during walking \cite{gandemer2017spatial, mahmud2022auditory}. Humans naturally orient toward sound sources, which may require less cognitive processing than interpreting vibrotactile spatial signals. A similar pattern appears in the bimodal comparison between Vibrotactile+Visual and Auditory+Visual. Both conditions share the same visual anchor, but pairing the visual cue with auditory feedback again resulted in better efficiency. This suggests that auditory signals complement visual cues more effectively during VR locomotion than vibrotactile signals in the current implementation. It should be noted, however, that this domination result is specific to the implementations used here.

Among the five frontier conditions, the efficiency ratio (gap closure per unit of burden) provides a useful way to compare options. \textbf{Spatial Auditory} achieved the highest efficiency ratio (0.086), making it a strong default option when minimizing user burden is important. In contrast, \textbf{Multimodal} feedback produced the largest performance improvement (95.6\% gap closure) but also imposed the highest burden. The remaining frontier conditions -- \textbf{Static Visual}, \textbf{Auditory+Visual}, and \textbf{Auditory+Vibrotactile} offer intermediate trade-offs that may be appropriate for different stages of rehabilitation.
\vspace{-0.1in}

\subsection{Robustness of the Efficiency Structure}
The complete stability of the Pareto frontier across all three burden operationalizations confirms \textbf{H5}. It addresses the most critical methodological concern about the framework that the frontier might be an artifact of how burden was measured rather than a genuine property of the data. But the same five conditions were dominant and the same two were dominated whether burden was the composite, mental load alone, or fatigue alone, demonstrating that the efficiency structure is invariant to how burden was calculated.
This has two implications. First, it supports the use of the burden composite, since both mental load and fatigue provide consistent information and combining them does not change the results. Second, it shows that the framework can work with simple measurements. Even basic Likert-scale ratings are enough to identify the same optimal conditions, making the approach practical for real-world settings where long assessments are not feasible.

\subsection{Limitations and Future Directions}
Several limitations should be considered when interpreting these results. First, the normative threshold $T$ was derived from a dataset of 41 MS participants with a small and non-significant difference between fallers and non-fallers ($d = 0.44$, $p = .16$). Although the non-faller mean provides a useful directional benchmark, larger datasets with confirmed fall outcomes are needed to validate $T$ as an individual rehabilitation target. Second, the burden composite was based on two single-item Likert scales measuring mental load and fatigue. While sensitivity analysis showed that the Pareto frontier remained stable across both measures, more detailed instruments such as NASA-TLX \cite{hart2006nasa} or physiological indicators (heart rate variability or EEG measures of cognitive load) could provide a more comprehensive assessment of user burden. Third, all measurements were collected within a single experimental session. Therefore, the results reflect short-term responses rather than long-term rehabilitation outcomes. Future longitudinal studies are needed to determine whether repeated exposure to the frontier conditions leads to sustained improvements in real-world walking.

As a part of future research, the framework was applied only to participants with MS and whether a similar Pareto structure would emerge in other populations, such as healthy older adults or individuals with Parkinson’s disease or stroke, remains an open question.
Finally, the visual feedback used in this study relied on a head-fixed anchor, which may be less effective than world-fixed alternatives \cite{shahnewaz2021static}. Future work should examine whether alternative visual feedback designs change the efficiency relationships observed here.
Despite these limitations, the threshold-anchored efficiency framework can be applied beyond MS and walking velocity. Similar approaches could be used to study balance in vestibular disorders, upper-limb movement in stroke rehabilitation, or other clinical tasks where both performance and user burden are relevant. The efficiency ratio introduced in Table~\ref{tab:pareto} may also support adaptive VR systems that dynamically adjust feedback conditions based on a patient’s current burden level.

\section{Conclusion}
This paper introduced a \emph{threshold-anchored pareto efficiency framework} for VR gait feedback selection for individuals with MS. We evaluated eight feedback conditions on two criteria simultaneously: \textbf{(a)} how much each condition closes the velocity gap to a normative fall-risk threshold ($T=1.29 m/s$) \textbf{(b)} what is the cognitive and physical cost. Three key findings emerged. Firstly, walking in VR without feedback reduced walking velocity by approx. 0.21 m/s compared to real-world walking, placing participants further below the threshold. Secondly, all eight feedback conditions improved walking relative to the VR baseline (closing 22.7\% to 95.6\%) of the velocity gap and improving gait performance. Lastly, performance gain and user burden were not directly aligned. Pareto efficiency analysis identified \emph{five} frontier conditions and \emph{two} dominated conditions. Among frontier conditions, \emph{Spatial Auditory} feedback showed highest efficiency closing 72.7\% of the velocity gap while maintaining below-average burden. As such, this framework provides a practical method for evaluating VR rehabilitation conditions and guiding clinically meaningful protocol design for people with MS.
\acknowledgments{

}

\bibliographystyle{abbrv-doi}

\bibliography{biblio}

@article{walton2020ms,
  author  = {Walton, Clare and King, Rachel and Rechtman, Lindsay
             and Kaye, Wendy and Leray, Emmanuelle and Marrie, Ruth Ann
             and Robertson, Neil and La Rocca, Nicholas and Uitdehaag, Bernard
             and van der Mei, Ingrid and Baneke, Pieter and Fully, Christine
             and Engelhard, Marc},
  title   = {Rising prevalence of multiple sclerosis worldwide:
             Insights from the {Atlas of MS}, third edition},
  journal = {Multiple Sclerosis Journal},
  volume  = {26},
  number  = {14},
  pages   = {1816--1821},
  year    = {2020},
  doi     = {10.1177/1352458520970841}
}

@article{motl2012exercise,
  author  = {Motl, Robert W. and Pilutti, Lara A.},
  title   = {The benefits of exercise training in multiple sclerosis},
  journal = {Nature Reviews Neurology},
  volume  = {8},
  number  = {9},
  pages   = {487--497},
  year    = {2012},
  doi     = {10.1038/nrneurol.2012.117}
}

@article{nilsagard2009falls,
  author  = {Nilsag{\aa}rd, Ylva and Lundholm, Claes and Denison, Eva
             and Gunnarsson, Lena-Pia},
  title   = {Predicting accidental falls in people with multiple sclerosis
             --- a longitudinal study},
  journal = {Clinical Rehabilitation},
  volume  = {23},
  number  = {3},
  pages   = {259--269},
  year    = {2009},
  doi     = {10.1177/0269215508095960}
}

@article{sosnoff2011falls,
  author  = {Sosnoff, Jacob J. and Socie, Michael J. and Boes, Morgan K.
             and Sandroff, Brian M. and Pula, John H. and Suh, Yoojin
             and Motl, Robert W.},
  title   = {Mobility, balance and falls in persons with multiple sclerosis},
  journal = {PLOS ONE},
  volume  = {6},
  number  = {11},
  pages   = {e28021},
  year    = {2011},
  doi     = {10.1371/journal.pone.0028021}
}

@article{comber2017gait,
  title={Gait deficits in people with multiple sclerosis: A systematic review and meta-analysis},
  author={Comber, Laura and Galvin, Rose and Coote, Susan},
  journal={Gait \& posture},
  volume={51},
  pages={25--35},
  year={2017},
  publisher={Elsevier}
}

@article{kalron2014relationship,
  title={The relationship between fear of falling to spatiotemporal gait parameters measured by an instrumented treadmill in people with multiple sclerosis},
  author={Kalron, Alon and Achiron, Anat},
  journal={Gait \& posture},
  volume={39},
  number={2},
  pages={739--744},
  year={2014},
  publisher={Elsevier}
}

@article{laver2017vr,
  author  = {Laver, Kate E. and Lange, Belinda and George, Susan
             and Deutsch, Judith E. and Saposnik, Gustavo and Crotty, Maria},
  title   = {Virtual reality for stroke rehabilitation},
  journal = {Cochrane Database of Systematic Reviews},
  volume  = {11},
  pages   = {CD008349},
  year    = {2017},
  doi     = {10.1002/14651858.CD008349.pub4}
}

@article{massetti2016virtual,
  title={Virtual reality in multiple sclerosis--a systematic review},
  author={Massetti, Thais and Trevizan, Isabela Lopes and Arab, Claudia and Favero, Francis Meire and Ribeiro-Papa, Denise Cardoso and de Mello Monteiro, Carlos Bandeira},
  journal={Multiple sclerosis and related disorders},
  volume={8},
  pages={107--112},
  year={2016},
  publisher={Elsevier}
}

@article{sigrist2013augmented,
  title={Augmented visual, auditory, haptic, and multimodal feedback in motor learning: a review},
  author={Sigrist, Roland and Rauter, Georg and Riener, Robert and Wolf, Peter},
  journal={Psychonomic bulletin \& review},
  volume={20},
  number={1},
  pages={21--53},
  year={2013},
  publisher={Springer}
}

@article{mirelman2010vr,
  author  = {Mirelman, Anat and Patritti, Beth L. and Bonato, Paolo
             and Deutsch, Judith E.},
  title   = {Effects of virtual reality training on gait biomechanics
             of individuals post-stroke},
  journal = {Gait \& Posture},
  volume  = {31},
  number  = {4},
  pages   = {433--437},
  year    = {2010},
  doi     = {10.1016/j.gaitpost.2010.01.016}
}

@article{calabro2017role,
  title={The role of virtual reality in improving motor performance as revealed by EEG: a randomized clinical trial},
  author={Calabr{\`o}, Rocco Salvatore and Naro, Antonino and Russo, Margherita and Leo, Antonino and De Luca, Rosaria and Balletta, Tina and Buda, Antonio and La Rosa, Gianluca and Bramanti, Alessia and Bramanti, Placido},
  journal={Journal of neuroengineering and rehabilitation},
  volume={14},
  number={1},
  pages={53},
  year={2017},
  publisher={Springer}
}

@techreport{iso9241,
  author      = {{International Organization for Standardization}},
  title       = {{ISO} 9241-11: Ergonomics of Human-System Interaction ---
                 {Part} 11: Usability: Definitions and Concepts},
  institution = {ISO},
  year        = {2018},
  number      = {ISO 9241-11:2018}
}

@article{meyer2022open,
  title={Open-source dataset reveals relationship between walking bout duration and fall risk classification performance in persons with multiple sclerosis},
  author={Meyer, Brett M and Tulipani, Lindsey J and Gurchiek, Reed D and Allen, Dakota A and Solomon, Andrew J and Cheney, Nick and McGinnis, Ryan S},
  journal={PLOS digital health},
  volume={1},
  number={10},
  pages={e0000120},
  year={2022},
  publisher={Public Library of Science San Francisco, CA USA}
}

@inproceedings{mahmud2022auditory,
  title={Auditory feedback to make walking in virtual reality more accessible},
  author={Mahmud, M Rasel and Stewart, Michael and Cordova, Alberto and Quarles, John},
  booktitle={2022 IEEE international symposium on mixed and augmented reality (ISMAR)},
  pages={847--856},
  year={2022},
  organization={IEEE}
}

@inproceedings{samaraweera2013latency,
  title={Latency and avatars in virtual environments and the effects on gait for persons with mobility impairments},
  author={Samaraweera, Gayani and Guo, Rongkai and Quarles, John},
  booktitle={2013 IEEE Symposium on 3D User Interfaces (3DUI)},
  pages={23--30},
  year={2013},
  organization={IEEE}
}

@inproceedings{mahmud2023eyes,
  title={The Eyes Have It: Visual Feedback Methods to Make Walking in Immersive Virtual Reality More Accessible for People With Mobility Impairments While Utilizing Head-Mounted Displays},
  author={Mahmud, M Rasel and Cordova, Alberto and Quarles, John},
  booktitle={Proceedings of the 25th International ACM SIGACCESS Conference on Computers and Accessibility},
  pages={1--10},
  year={2023}
}

@article{kim2023virtual,
  title={Virtual reality-based gait rehabilitation intervention for stroke individuals: a scoping review},
  author={Kim, Minjoon and Kaneko, Fuminari},
  journal={Journal of exercise rehabilitation},
  volume={19},
  number={2},
  pages={95},
  year={2023}
}

@article{keshner2021untapped,
  title={The untapped potential of virtual reality in rehabilitation of balance and gait in neurological disorders},
  author={Keshner, Emily A and Lamontagne, Anouk},
  journal={Frontiers in virtual reality},
  volume={2},
  pages={641650},
  year={2021},
  publisher={Frontiers Media SA}
}

@manual{gaitrite_manual,
  title   = {{GaitRite} Manual: Technical aspects and gait parameters for the GAITRite Walkway System},
  author  = {{ProCare BV}},
  year    = {2017},
  url     = {https://www.procarebv.nl/wp-content/uploads/2017/01/Technische-aspecten-GAITrite-Walkway-System.pdf},
  urldate = {2025-03-30},
  note    = {PDF (Technical aspects of the GAITRite Walkway System)}
}

@article{mahmud2024multimodal,
  title={Multimodal feedback methods for advancing the accessibility of immersive virtual reality for people with balance impairments due to multiple sclerosis},
  author={Mahmud, M Rasel and Cordova, Alberto and Quarles, John},
  journal={IEEE Transactions on Visualization and Computer Graphics},
  year={2024},
  publisher={IEEE}
}

@inproceedings{mahmud2023auditory,
  title={Auditory, Vibrotactile, or Visual? Investigating the Effective Feedback Modalities to Improve Standing Balance in Immersive Virtual Reality for People with Balance Impairments Due to Type 2 Diabetes},
  author={Mahmud, M Rasel and Cordova, Alberto and Quarles, John},
  booktitle={2023 IEEE International Symposium on Mixed and Augmented Reality (ISMAR)},
  pages={573--582},
  year={2023},
  organization={IEEE}
}

@article{helps2014different,
  title={Different effects of adding white noise on cognitive performance of sub-, normal and super-attentive school children},
  author={Helps, Suzannah K and Bamford, Susan and Sonuga-Barke, Edmund JS and S{\"o}derlund, G{\"o}ran BW},
  journal={PloS one},
  volume={9},
  number={11},
  pages={e112768},
  year={2014},
  publisher={Public Library of Science San Francisco, USA}
}

@inproceedings{mahmud2025vibrotactile,
  title={Vibrotactile Feedback to Make Real Walking in Virtual Reality More Accessible for People With and Without Mobility Impairments},
  author={Mahmud, M Rasel and Stewart, Michael and Cordova, Alberto and Quarles, John},
  booktitle={Proceedings of the 2025 31st ACM Symposium on Virtual Reality Software and Technology},
  pages={1--2},
  year={2025}
}

@inproceedings{mahmud2022standing,
  title={Standing balance improvement using vibrotactile feedback in virtual reality},
  author={Mahmud, M Rasel and Stewart, Michael and Cordova, Alberto and Quarles, John},
  booktitle={Proceedings of the 28th ACM Symposium on Virtual Reality Software and Technology},
  pages={1--11},
  year={2022}
}

@article{shahnewaz2021static,
  title={Static rest frame to improve postural stability in virtual and augmented reality},
  author={Shahnewaz Ferdous, Sharif Mohammad and Chowdhury, Tanvir Irfan and Arafat, Imtiaz Muhammad and Quarles, John},
  journal={Frontiers in Virtual Reality},
  volume={1},
  pages={582169},
  year={2021},
  publisher={Frontiers Media SA}
}

@article{kennedy1993simulator,
  title={Simulator sickness questionnaire: An enhanced method for quantifying simulator sickness},
  author={Kennedy, Robert S and Lane, Norman E and Berbaum, Kevin S and Lilienthal, Michael G},
  journal={The international journal of aviation psychology},
  volume={3},
  number={3},
  pages={203--220},
  year={1993},
  publisher={Taylor \& Francis}
}

@article{powell1995activities,
  title={The activities-specific balance confidence (ABC) scale},
  author={Powell, Lynda Elaine and Myers, Anita M},
  journal={The journals of Gerontology Series A: Biological sciences and Medical sciences},
  volume={50},
  number={1},
  pages={M28--M34},
  year={1995},
  publisher={The Gerontological Society of America}
}

@inproceedings{bouchard2004reliability,
  title={Reliability and validity of a single-item measure of presence in VR},
  author={Bouchard, St{\'e}phane and Robillard, Genevi{\'e}ve and St-Jacques, Julie and Dumoulin, St{\'e}phanie and Patry, Marie-Jos{\'e}e and Renaud, Patrice},
  booktitle={The 3rd IEEE international workshop on haptic, audio and visual environments and their applications},
  pages={59--61},
  year={2004},
  organization={IEEE}
}

@article{horsak2021overground,
  title={Overground walking in a fully immersive virtual reality: A comprehensive study on the effects on full-body walking biomechanics},
  author={Horsak, Brian and Simonlehner, Mark and Sch{\"o}ffer, Lucas and Dumphart, Bernhard and Jalaeefar, Arian and Husinsky, Matthias},
  journal={Frontiers in bioengineering and biotechnology},
  volume={9},
  pages={780314},
  year={2021},
  publisher={Frontiers Media SA}
}

@inproceedings{canessa2019comparing,
  title={Comparing Real Walking in Immersive Virtual Reality and in Physical World using Gait Analysis.},
  author={Canessa, Andrea and Casu, Paolo and Solari, Fabio and Chessa, Manuela},
  booktitle={VISIGRAPP (2: HUCAPP)},
  pages={121--128},
  year={2019}
}

@inproceedings{ferdous2018investigating,
  title={Investigating the reason for increased postural instability in virtual reality for persons with balance impairments},
  author={Ferdous, Sharif Mohammad Shahnewaz and Chowdhury, Tanvir Irfan and Arafat, Imtiaz Muhammad and Quarles, John},
  booktitle={Proceedings of the 24th ACM Symposium on Virtual Reality Software and Technology},
  pages={1--7},
  year={2018}
}

@article{asslander2014sensory,
  title={Sensory reweighting dynamics in human postural control},
  author={Assl{\"a}nder, Lorenz and Peterka, Robert J},
  journal={Journal of neurophysiology},
  volume={111},
  number={9},
  pages={1852--1864},
  year={2014},
  publisher={American Physiological Society Bethesda, MD}
}

@article{roy2021multisensory,
  title={Multisensory integration and behavioral stability},
  author={Roy, Charlotte and Dalla Bella, Simone and Pla, Simon and Lagarde, Julien},
  journal={Psychological Research},
  volume={85},
  number={2},
  pages={879--886},
  year={2021},
  publisher={Springer}
}

@article{wall2010application,
  title={Application of vibrotactile feedback of body motion to improve rehabilitation in individuals with imbalance},
  author={Wall III, Conrad},
  journal={Journal of Neurologic Physical Therapy},
  volume={34},
  number={2},
  pages={98--104},
  year={2010},
  publisher={LWW}
}

@article{gandemer2017spatial,
  title={Spatial cues provided by sound improve postural stabilization: evidence of a spatial auditory map?},
  author={Gandemer, Lennie and Parseihian, Gaetan and Kronland-Martinet, Richard and Bourdin, Christophe},
  journal={Frontiers in neuroscience},
  volume={11},
  pages={357},
  year={2017},
  publisher={Frontiers Media SA}
}

@article{askari2021cognitive,
  title={Cognitive fatigue interventions for people with multiple sclerosis: A scoping review},
  author={Askari, Sorayya and Fanelli, Domenica and Harvey, Keri},
  journal={Multiple Sclerosis and Related Disorders},
  volume={55},
  pages={103213},
  year={2021},
  publisher={Elsevier}
}

@inproceedings{hart2006nasa,
  title={NASA-task load index (NASA-TLX); 20 years later},
  author={Hart, Sandra G},
  booktitle={Proceedings of the human factors and ergonomics society annual meeting},
  volume={50},
  pages={904--908},
  year={2006},
  organization={Sage publications Sage CA: Los Angeles, CA}
}

@article{d2012cognitive,
  title={Cognitive and motor functioning in patients with multiple sclerosis: neuropsychological predictors of walking speed and falls},
  author={D'Orio, Vanessa L and Foley, Frederick W and Armentano, Francine and Picone, Mary Ann and Kim, Sonya and Holtzer, Roee},
  journal={Journal of the neurological sciences},
  volume={316},
  number={1-2},
  pages={42--46},
  year={2012},
  publisher={Elsevier}
}

@article{motl2017validity,
  title={Validity of the timed 25-foot walk as an ambulatory performance outcome measure for multiple sclerosis},
  author={Motl, Robert W and Cohen, Jeffrey A and Benedict, Ralph and Phillips, Glenn and LaRocca, Nicholas and Hudson, Lynn D and Rudick, Richard and Multiple Sclerosis Outcome Assessments Consortium},
  journal={Multiple Sclerosis Journal},
  volume={23},
  number={5},
  pages={704--710},
  year={2017},
  publisher={SAGE Publications Sage UK: London, England}
}

@article{janeh2019gait,
  title={Gait training in virtual reality: short-term effects of different virtual manipulation techniques in Parkinson’s disease},
  author={Janeh, Omar and Fr{\"u}ndt, Odette and Sch{\"o}nwald, Beate and Gulberti, Alessandro and Buhmann, Carsten and Gerloff, Christian and Steinicke, Frank and P{\"o}tter-Nerger, Monika},
  journal={Cells},
  volume={8},
  number={5},
  pages={419},
  year={2019},
  publisher={MDPI}
}

@article{winter2021immersive,
  title={Immersive virtual reality during gait rehabilitation increases walking speed and motivation: a usability evaluation with healthy participants and patients with multiple sclerosis and stroke},
  author={Winter, Carla and Kern, Florian and Gall, Dominik and Latoschik, Marc Erich and Pauli, Paul and K{\"a}thner, Ivo},
  journal={Journal of neuroengineering and rehabilitation},
  volume={18},
  number={1},
  pages={68},
  year={2021},
  publisher={Springer}
}

@article{kingma2019vibrotactile,
  title={Vibrotactile feedback improves balance and mobility in patients with severe bilateral vestibular loss},
  author={Kingma, Herman and Felipe, Lilian and Gerards, Marie-Cecile and Gerits, Peter and Guinand, Nils and Perez-Fornos, Angelica and Demkin, Vladimir and Van De Berg, Raymond},
  journal={Journal of neurology},
  volume={266},
  number={Suppl 1},
  pages={19--26},
  year={2019},
  publisher={Springer}
}

@article{guo2015mobility,
  title={Mobility impaired users respond differently than healthy users in virtual environments},
  author={Guo, Rongkai and Samaraweera, Gayani and Quarles, John},
  journal={Computer Animation and Virtual Worlds},
  volume={26},
  number={5},
  pages={509--526},
  year={2015},
  publisher={Wiley Online Library}
}

@article{lee2015influence,
  title={Influence of visual feedback training on the balance and walking in stroke patients},
  author={Lee, Kwan-Sub and Choe, Han-Seong and Lee, Jae-Hong},
  journal={The Journal of Korean Physical Therapy},
  volume={27},
  number={6},
  pages={407--412},
  year={2015},
  publisher={The Korea Society of Physical Therapy}
}

@article{weller2022redirected,
  title={Redirected walking in virtual reality with auditory step feedback},
  author={Weller, Rene and Brennecke, Benjamin and Zachmann, Gabriel},
  journal={The Visual Computer},
  volume={38},
  number={9},
  pages={3475--3486},
  year={2022},
  publisher={Springer}
}

@article{machado2023novel,
  title={A novel mixed reality assistive system to aid the visually and mobility impaired using a multimodal feedback system},
  author={Machado, Fabiana and Loureiro, Matheus and Mello, Ricardo C and Diaz, Camilo AR and Frizera, Anselmo},
  journal={Displays},
  volume={79},
  pages={102480},
  year={2023},
  publisher={Elsevier}
}

@article{leonardis2014multisensory,
  title={Multisensory feedback can enhance embodiment within an enriched virtual walking scenario},
  author={Leonardis, Daniele and Frisoli, Antonio and Barsotti, Michele and Carrozzino, Marcello and Bergamasco, Massimo},
  journal={Presence},
  volume={23},
  number={3},
  pages={253--266},
  year={2014},
  publisher={MIT Press}
}

@article{kwakkel2004understanding,
  title={Understanding the pattern of functional recovery after stroke: facts and theories},
  author={Kwakkel, Gert and Kollen, Boudewijn and Lindeman, Eline},
  journal={Restorative neurology and neuroscience},
  volume={22},
  number={3-5},
  pages={281--299},
  year={2004},
  publisher={SAGE Publications Sage UK: London, England}
}

@incollection{deb2011multi,
  title={Multi-objective optimisation using evolutionary algorithms: an introduction},
  author={Deb, Kalyanmoy},
  booktitle={Multi-objective evolutionary optimisation for product design and manufacturing},
  pages={3--34},
  year={2011},
  publisher={Springer}
}

@article{induruwa2012fatigue,
  title={Fatigue in multiple sclerosis—a brief review},
  author={Induruwa, Isuru and Constantinescu, Cris S and Gran, Bruno},
  journal={Journal of the neurological sciences},
  volume={323},
  number={1-2},
  pages={9--15},
  year={2012},
  publisher={Elsevier}
}

@article{chang2020virtual,
  title={Virtual reality sickness: a review of causes and measurements},
  author={Chang, Eunhee and Kim, Hyun Taek and Yoo, Byounghyun},
  journal={International Journal of Human--Computer Interaction},
  volume={36},
  number={17},
  pages={1658--1682},
  year={2020},
  publisher={Taylor \& Francis}
}

@article{de2016effect,
  title={Effect of virtual reality training on balance and gait ability in patients with stroke: systematic review and meta-analysis},
  author={De Rooij, Ilona JM and Van De Port, Ingrid GL and Meijer, Jan-Willem G},
  journal={Physical therapy},
  volume={96},
  number={12},
  pages={1905--1918},
  year={2016},
  publisher={Oxford University Press}
}

@article{feng2019virtual,
  title={Virtual reality rehabilitation versus conventional physical therapy for improving balance and gait in Parkinson’s disease patients: a randomized controlled trial},
  author={Feng, Hao and Li, Cuiyun and Liu, Jiayu and Wang, Liang and Ma, Jing and Li, Guanglei and Gan, Lu and Shang, Xiaoying and Wu, Zhixuan},
  journal={Medical science monitor: international medical journal of experimental and clinical research},
  volume={25},
  pages={4186},
  year={2019}
}
\end{document}